\documentclass[aps,pra,twocolumn,superscriptaddress,showpacs,floatfix]{revtex4}
\usepackage{graphicx,dcolumn,bm,hyperref,amsmath,amssymb,xspace,epsfig,float,array,multirow,amsfonts}
\usepackage[vcentermath]{youngtab}
\usepackage{relsize,exscale}
\usepackage{amsmath}
\usepackage{graphicx}
\usepackage{gensymb}
\usepackage[colorinlistoftodos]{todonotes}
\usepackage{color}
\usepackage[english]{babel}
\usepackage[utf8]{inputenc}
\usepackage{tikz,pgfplots}
\usepackage{amsmath}
\usepackage{graphicx}
\usepackage{wasysym}
\usepackage[colorinlistoftodos]{todonotes}
\usepackage{mathtools, bm}
\usepackage{amssymb, bm}
\usepackage{amsmath}

\newcommand{\subfigimg}[3][,]{%
		\setbox1=\hbox{\includegraphics[#1]{#3}}
		\leavevmode\rlap{\usebox1}
		\rlap{\hspace*{2pt}\raisebox{\dimexpr\ht1-0.5\baselineskip}{{\bfseries \large\textsf{#2}}}}
		\phantom{\usebox1}
}

\begin{document}

\title{Excitation spectrum and supersolidity of a two-leg bosonic ring ladder}

\author{Nicolas Victorin}
\affiliation{Univ.~Grenoble Alpes, CNRS, LPMMC, 38000 Grenoble France}

\author{Paolo Pedri}
\affiliation{Laboratoire de Physique des Lasers, CNRS, Universit\'e Paris 13, Sorbonne Paris Cit\'e, 99 avenue J.-B. Cl\'ement, F-93430 Villetaneuse, France}

\author{Anna Minguzzi}
\affiliation{Univ.~Grenoble Alpes, CNRS, LPMMC, 38000 Grenoble France}

\date{\today}

\begin{abstract}
  We consider a system of weakly interacting bosons confined on a planar double lattice ring subjected to two artificial gauge fields. This system is known to display three phases,  the Meissner phase  where the flow of particles is carried at the edges of the system without transverse current, a vortex phase characterized by non-zero transverse current, and a  biased-ladder phase, characterized by an imbalance of the population of the two rings. We use the Bogoliubov approximation to determine the excitation spectrum in the three phases, the dynamic structure factor and the quantum fluctuation corrections to the first-order correlation function. Our analysis reveals supersolid features as well as Josephson modes, corresponding to out-of-phase modes of the finite ring.
\end{abstract}


\maketitle

\maketitle
\section{Introduction}
Supersolidity is a combined effect of solid order and superfluid flow. In a bosonic system, a supersolid may be formed by  breaking two symmetries: a) the continuous  translational symmetry  in order to create a crystal order  (or discrete translational symmetry on a lattice) and b) the $U(1)$ symmetry in order to create a Bose-Einstein condensate. The latter  is a superfluid thanks to irrotationality of the velocity field associated to the condensate wavefunction. The concept of supersolidity was first introduced in the context of liquid Helium more than fifty years ago~\cite{LegettSuper,Chester}. With ultracold atoms,  supersolidity has been observed with  Bose-Einstein condensates in a cavity ~\cite{DickeQuantumPhase,Leonard2017a,Leonard2017b,Li2017} as well as in dipolar quantum gases \cite{Boettcher2019,Tanzi2019,Chomaz2019,Natale2019}.
Both experimental realizations of supersolidity are based on long-range interactions~\cite{Quantumphasesfromcompetingshort-andlong-rangeinteractions},  a key ingredient first introduced by Gross~\cite{Gross}. Another route to reach supersolidity is to have a peculiar single-particle dispersion, eg one with two degenerate minima.  An example of this second case is provided by spin-orbit coupled Bose gases where supersolidity has also been studied ~\cite{Spinorbit,SpinoExp}. Experimentally, crystal order in spin-orbit coupled Bose gases has been evidenced by the observation of stripes~{\cite{Ketterle}. However, the visibility of the fringes of the density  is limited  by interspecies interactions and  is a major issue to overcome. 

In this work, we consider a two-leg bosonic ring lattice subjected to two gauge fields.  Here, thanks to the peculiar geometry of the system the inter-species interactions can be completely suppressed hence providing a new arena for studying  supersolidity in a condition of high fringe visibility. As for the case of a spin-orbit coupled Bose gas,  in a two-leg bosonic ring ladder there is  no explicit long-range interaction but it emerges as an effective low-energy property due to the effect of gauge field and tunnel coupling between the rings.

The  bosonic ladder under a  gauge field in a linear geometry has been the object of intense theoretical work,  by means of DMRG simulations~\cite{PhysRevB.91.140406DMRG} and field theoretical methods~\cite{Georges,PhysRevB.64.144515Orignac}.    Those studies have provided  a complete  characterization of the phase diagram of this system, showing various phase.  Among them, we mention the chiral superfluid phases, the chiral Mott insulating phases displaying Meissner currents ~\cite{PhysRevB.64.144515Orignac,ChiralMottDhar} and vortex-Mott insulating phases~\cite{PhysRevLett.111.150601}. In the weakly interacting regime, on which we will focus on this work, an additional phase has been predicted ~\cite{Mueller}: the biased-ladder phase, characterized by an imbalanced population of the bosons between the two legs, explicitly breaking $\mathbb{Z}_2$ symmetry. In parallel to these theoretical advances, the experimental realization of the bosonic flux ladder has been reported in optical lattices~\cite{Atala} as well as for lattices in  synthetic dimensions, both for fermionic and bosonic quantum gases~\cite{Stuhl,Mancini}.

In this paper we provide several indications for supersolid features of the double-ring ladder system  subjected to different flux in each leg. First of all, we study the properties of the excitation spectrum in order to demonstrate  the first-order coherence. In the Meissner phase  we find  a single Goldstone mode, associated to Bose-Einstein condensation in the only  minimum of the single-particle dispersion relation.
In the biased ladder phase, in addition to the phonon branch we predict the existence of a roton minimum. This is the precursor of the vortex phase,  in agreement with previews studies \cite{Mueller}. In the vortex phase, two Goldstone modes are observed, associated to the spontaneous breaking of $U(1)$ and translational symmetry. A further proof of the coherence properties of the system is provided by the calculation of the first-order spatial correlation function.

As a second step, we then provide various indications of spatial crystal-like  order. First of all, in the excitation spectrum of the vortex phases we find a   folding of the Brillouin zone. This is a consequence of  the formation of spatial modulations  in the mean-field condensate density due to the formation of a vortex lattice along the ring.  Furthermore, the analysis of the static structure factor shows the emergence of a peak at finite wavevector, corresponding to the density modulation along the rings.

Combining together the various evidences of coherence properties and crystalline order, we obtain a univocal evidence of supersolidity.  Coupled rings under gauge fields hence provide a novel platform for the experimental study of supersolid order with ultracold atoms.

Finally, we address some features peculiar to the finite ring case, and  in particular the emergence of Josephson  modes for weakly coupled rings.  These modes correspond to  a spatially uniform out-of-phase oscillation of particles between the rings, hence providing a further indication of the coherence among the two rings.

\section{Model and method}
We consider a Bose gas confined in a double ring lattice (see Fig.~\ref{fig1b}). In the tight-binding approximation we model the system using the Bose-Hubbard Hamiltonian:
\begin{align}
&\hat{H}=\hat{H}_0+\hat{H}_{int}=\nonumber\\ &-\!\sum_{l=1,p=1,2}^{N_s} \!\!\!J_p\left(\hat{a}^{\dagger}_{l,p} \hat{a}_{l+1,p}e^{i\Phi_p} + \hat{a}^{\dagger}_{l+1,p}\hat{a}_{l,p}e^{-i\Phi_p}\right)\nonumber \\ &-\!K\sum_{l=1}^{N_s}\left(\hat{a}_{l,1}^{\dagger}\hat{a}_{l,2}+\hat{a}_{l,2}^{\dagger}\hat{a}_{l,1}\right)+\frac{U}{2}\!\!\!\sum_{l=1,p=1,2}^{N_s}a^{\dagger}_{l,p}\hat{a}^{\dagger}_{l,p}\hat{a}_{l,p}\hat{a}_{l,p}
\label{eq1}
\end{align}
where the position of a particle on each ring is indicated by an integer $l \in \left[1, N_s\right]$ with $N_s$ the number of sites in each ring, and $\hat{a}_{l,p}$,  $\hat{a}_{l,p}^\dagger$ are respectively the bosonic destruction and creation operators at site $l$ each ring, with $p=1,2$ indicting the inner and outer ring respectively. 
In Eq.~(\ref{eq1}) $J_1$ and $J_2$ are  the tunneling amplitudes from one site to another along each ring,  $K$ is the tunneling amplitude between the two rings, connecting only sites with the same position index $l$ and $\Phi_{1,2}$ are the fluxes threading each ring.

\begin{figure}[t]
\centerline{\includegraphics[scale=0.4]{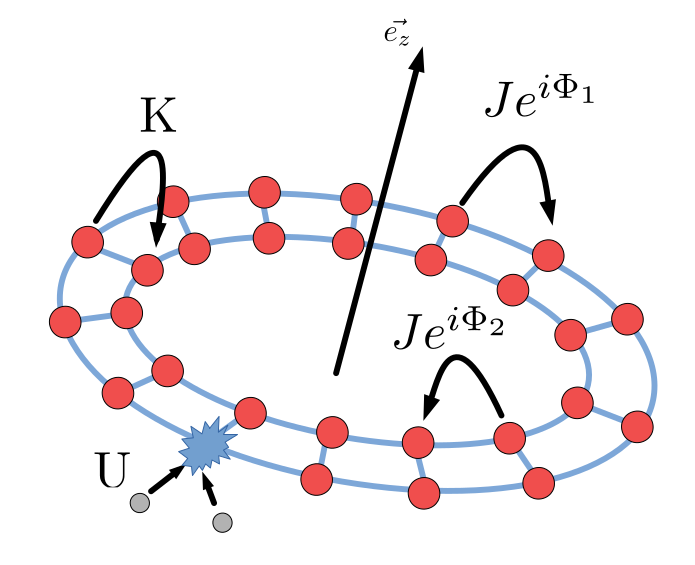}}
\caption{Sketch of the system studied in this work: coplanar ring lattices with the same number of sites, with inter-ring tunnel energy $K$, intra-ring tunnel energies $Je^{i\Phi_p}$ with $p=1,2$ and on-site  interaction  $U$.}
\label{fig1b}
\end{figure}

\subsection{Bogoliubov De-Gennes equations }

In order to obtain the excitation spectrum of the double ring in the Bogoliubov approximation,  we start from the Heisenberg equations of motion for the bosonic field operators $\hat{a}_{l,\alpha}$ ~\cite{Fetter} and we replace $\hat{a}_{l,\alpha}$ in the quantum Hamiltonian (\ref{eq1}) by $\hat{a}_{l,\alpha}=\Psi^{(0)}_{l,\alpha}+\delta\hat{a}_{l,\alpha}$. The field $\Psi^{(0)}_{l,\alpha}$ is the ground-state
condensate wave function, solution of the coupled discrete non linear Schrödinger equations (\ref{dnlse}) 
\begin{eqnarray}
\label{dnlse}
\mu  \Psi_{l,1}&=& -J\Psi_{l+1,1}e^{i(\Phi+\phi/2)}-J\Psi_{l-1,1}e^{-i(\Phi+\phi/2)}\nonumber\\&-&K\Psi_{l,2}+U|\Psi_{l,1}|^2\Psi_{l,1}\nonumber\\
\mu \Psi_{l,2}&=& -J\Psi_{l+1,2}e^{i(\Phi-\phi/2)}-J\Psi_{l-1,2}e^{-i(\Phi-\phi/2)}\nonumber\\&-&K\Psi_{l,1}+U|\Psi_{l,2}|^2\Psi_{l,2},
\end{eqnarray}
where $\Phi=\frac{\Phi_1+\Phi_2}{2}$,  $\phi= \Phi_1-\Phi_2$ and $\mu$ is the chemical potential. In the following,  we will consider for simplicity the case $\Phi=0$. The next step consists in the expansion and truncation of the Hamiltonian  (\ref{eq1}) to quadractic order  in  $\delta\hat{a}_{l,p}$,   $\delta\hat{a}_{l,p}^\dagger$,  yielding the   Bogoliubov Hamiltonian 
\begin{eqnarray}
\hat{H}_{Bog}=\left(\delta\mathbf{\hat{a}^{\dagger}}_{1},\delta\mathbf{\hat{a}}_{1},\delta\mathbf{\hat{a}^{\dagger}}_{2},\delta\mathbf{\hat{a}}_{2}\right)H^{(2)}\begin{pmatrix}
\delta\mathbf{\hat{a}}_{1}\\
\delta\mathbf{\hat{a}}^{\dagger}_{1}\\
\delta\mathbf{\hat{a}}_{2}\\
\delta\mathbf{\hat{a}}^{\dagger}_{2}
\end{pmatrix},
\end{eqnarray}
where  $\delta \mathbf{\hat{a}^{\dagger}}_{p}=(\delta\hat{a}^{\dagger}_{1,p}, \delta\hat{a}^{\dagger}_{2,p}, ... ,\delta\hat{a}^{\dagger}_{l,p})  $ and the matrix $H^{(2)}$ can be written in the following form 
\begin{align}
H^{(2)}=\begin{pmatrix}
\mathbf{A}_1 & \mathbf{B}_1 & -\mathbf{K} & 0 \\
\mathbf{B}_1^*&\mathbf{A}_1^*&0&-\mathbf{K} \\
-\mathbf{K} &0&\mathbf{A}_2&\mathbf{B}_2 \\
0& -\mathbf{K}&\mathbf{B}_2^*&\mathbf{A}_2^*
\end{pmatrix},
\label{Hbog}
\end{align}
The matrix $\mathbf{A}_{p}$ with $p=1,2$ is given by
\begin{widetext}
\begin{equation}
\mathbf{A}_p=\left (
\begin{array}{ccc}
\begin{array}{ll}
2U|\Psi^{(0)}_{1,p}|^2 & -Je^{i\Phi_{p}} \\
-Je^{-i\Phi_{p}}& 2U|\Psi^{(0)}_{2,p}|^2
\end{array}
& \cdots & 
\begin{array}{l}
-Je^{-i\Phi_{p}}
\end{array} \\
\vdots & \ddots & \vdots\\
\begin{array}{ll}
-Je^{i\Phi_{p}} 
\end{array} &
\cdots & 
\begin{array}{ll}
2U|\Psi^{(0)}_{N_s-1,p}|^2 & -Je^{i\Phi_{p}} \\
-Je^{-i\Phi_{p}}& 2U|\Psi^{(0)}_{N_s,p}|^2
\end{array} 
\end{array}
\right),
\end{equation}
\end{widetext}
and $\mathbf{B}_{p}$ and $\mathbf{K}_{p}$ are diagonal matrices of dimension $N_s\times N_s$, i.e. $\mathbf{B}_p=$diag$(U(\Psi^{(0)}_{1,p})^2,...,U(\Psi^{(0)}_{N_s,p})^2)$,$\mathbf{K}=K\mathbf{I}$, with $\mathbf{I}$ the identity matrix.
We search then a transformation to quasi-particle operators $\hat{\gamma}_{\nu}$ for an excitation in mode $\nu$, such that the Bogoliubov Hamiltonian takes diagonal form:
\begin{align}
H_{Bog}=\sum_{\nu}\hbar\omega_{\nu}\hat{\gamma}_{\nu}^{\dagger}\hat{\gamma}_{\nu}
\end{align}
We use the following general transformation, where the operators $\hat{\gamma}_{\nu}$ follow usual bosonic commutation rules, $\left[\hat{\gamma}_{\nu},\hat{\gamma}_{\nu'}\right]=0$, $\left[\hat{\gamma}_{\nu},\hat{\gamma}^{\dagger}_{\nu'}\right]=\delta_{\nu,\nu'}$.
\begin{align}
\delta \hat{a}_{l,p} =\sum_{\nu} h_{\nu,l}^{(p)}\hat{\gamma}_{\nu}-Q_{\nu,l}^{*(p)}\hat{\gamma}_{\nu}^{\dagger},
\label{Transfo}
\end{align}
As next step, we substitute Eq.~(\ref{Transfo}) into the equation of motion, and use  the following properties
\begin{align}
&\left[\hat{\gamma}_{\nu},H\right]=\hbar\omega_{\nu}\hat{\gamma}_{\nu}\\
&\left[\hat{\gamma}_{\nu}^{\dagger},H\right]=-\hbar\omega_{\nu}\hat{\gamma}_{\nu}^{\dagger}.
\end{align}
Finally, by equating the coefficients of the differents modes  $\{h_{\nu}^{(p)},Q_{\nu}^{(p)}\}$ we obtain that the modes have to verify the following eigenvalue problem, corresponding to the Bogoliubov-De Gennes equations for the ring ladder:
\begin{widetext}
\begin{align}
\epsilon_{\nu} \begin{pmatrix}
\mathbf{h}_{\nu}^{(1)}\\
\mathbf{Q}_{\nu}^{(1)}\\
\mathbf{h}_{\nu}^{(2)}\\
\mathbf{Q}_{\nu}^{(2)}
\end{pmatrix}
=\begin{pmatrix}
\mathbf{A}_1-\mu\mathbb{\mathbf{I}} & \mathbf{B}_1 & -\mathbf{K} & 0 \\
-\mathbf{B}_1^*&-\mathbf{A}_1^*+\mu\mathbb{\mathbf{I}}&0&\mathbf{K} \\
-\mathbf{K} &0&\mathbf{A}_2-\mu\mathbb{\mathbf{I}}&\mathbf{B}_2 \\
0& \mathbf{K}&-\mathbf{B}_2^*&-\mathbf{A}_2^*+\mu\mathbb{\mathbf{I}}
\end{pmatrix}
\begin{pmatrix}
\mathbf{h}_{\nu}^{(1)}\\
\mathbf{Q}_{\nu}^{(1)}\\
\mathbf{h}_{\nu}^{(2)}\\
\mathbf{Q}_{\nu}^{(2)}
\end{pmatrix},
\label{BogDe}
\end{align}
\end{widetext}
where $\mathbf{h}_{\nu}^{(p)} = (h_{\nu,1}^{(p)}...h_{\nu,l}^{(p)} ...h_{\nu,N_s}^{(p)})^T$ and  $\mathbf{Q}_{\nu}^{(p)} = (Q_{\nu,1}^{(p)}...Q_{\nu,l}^{(p)} ...Q_{\nu,N_s}^{(\alpha)})^T$ and the chemical potential is $\mu=\langle\Psi_{(0)} |H_0|\Psi_{(0)} \rangle +2\langle \Psi_{(0)}|H_{int}|\Psi_{(0)}\rangle$ 
The eigenmodes satisfy the following orthogonality relations which follow from commutation relations among $\hat{\gamma}_{\nu}$:
\begin{align}
\sum_{\nu,p}h_{\nu,l}^{p}(h^{(p)}_{\nu,l'})^*-Q_{\nu,l}^{p}(Q^{(p)}_{\nu,l'})^*=\delta_{l,l'}
\label{biorth1}
\end{align}
\begin{align}
\sum_{l,p}h_{\nu,l}^{p}(h^{(p)}_{\nu',l})^*-Q_{\nu,l}^{p}(Q^{(p)}_{\nu',l})^*=\delta_{\nu,\nu'}.
\label{biorth2}
\end{align}

\subsection{Dynamical structure factor}

The dynamical structure factor is a powerful tool to study correlations in many-body systems both theoretically and experimentally. It corresponds to the space- and time- Fourier transform of the density-density correlation function.
The poles of the dynamical structure correspond to the collective excitation spectrum of the system. In the simplest case of a single-component one-dimensional system, the dynamical structure factor is defined as follows ~\cite{Dyna}
\begin{align}
S(q,\omega)=\sum_{s\neq 0} |\langle s |\hat \rho_q|0\rangle|^2\delta(\omega-\omega_s),
\end{align}
where  $q$ and $\omega$ are the momentum and energy transferred by the probe to the sample, $| s\rangle$ are many-body eigenstates of the system  and $|0\rangle$ is the ground state, $\hat{\rho}_q$ is the density fluctuation operator in momentum space and $\omega_s= E_s-E_0$ is the energy difference between excited and ground state.

For the case of coupled rings, since the excitations belong to both rings, we need to define several dynamical structure factors: $S_{p,p'}(q,\omega)$ with $p,p'=1,2$ being the ring index, and
\begin{align}
S_{p,p'}(q,\omega)=\sum_{s\neq 0}|\langle s |\hat \rho_{q}^{(p,p')}|0\rangle|^2\delta(\omega-\omega_s)
\end{align}
with
\begin{align}
\hat{\rho}_{q}^{(p,p')} = \sum_k \hat{a}_{k+q,p}^{\dagger}\hat{a}_{k,p'},
\end{align}
and $q$ and $k$ are wavevectors corresponding to the longitudinal momentum along each ring, ie we have set $\hat{a}_{k,p}=(1/\sqrt{N_s}) \sum_j \exp(i k j) \hat a_{j,p} $
Using the expansion (\ref{Transfo}) onto Bogoliubov modes, one can show that in the Bogoliubov approximation the dynamical structure factors are given by
\begin{align}
&S_{p,p'}(q,\omega)=\nonumber\\
&\sum_{s\neq 0}\left|\sum_{l}\left(\Psi_{l,p'}^{(0)}h_{s,l}^{*(p)}-\Psi_{l,p}^{*(0)}Q_{s,l}^{*(p')}\right)e^{iql}\right|^2\delta(\omega-\omega_s)
\label{Dynamical}
\end{align}

In order to understand the low energy properties of the system, instead of using the operators $\hat{a}_{k,p}$, it is useful to refer to the operators $\hat{\alpha}_k$ and $\hat{\beta}_k$ that diagonalize the single-particle non-interacting Hamiltonian $H_0$ (see Appendix A and Eq.~\ref{Diag} for its definition) and  introduce the strucure factors $S_{\lambda,\lambda'}$ where $\lambda,\lambda'=\alpha$ or $\beta$ referring to the corresponding operators. In particular, the low-energy properties of the system under study are governed by the lowest excitation branch associated to the operator $\hat{\beta}_k$ and can be accessed by studying the   dynamical structure factor  $S_{\beta,\beta}$: 
\begin{align}
S_{\beta,\beta}(q,\omega)=\sum_{s\neq 0}|\langle s|\hat \rho^{(\beta)}_q|0\rangle |^2 \delta(\omega-\omega_s)
\end{align}
where we have defined $\hat \rho^{\beta}_q =\sum_k\hat{\beta}_{k+q}^{\dagger}\hat{\beta}_k$. 

\subsection{Static structure factor}
The static structure factor yields information on spatial long-range order, eg crystal or density wave order, hence it is particularly suited to address the spatial modulations emerging in the vortex phase (see Sec.III below). The static structure factor is defined as
\begin{align}
S_{p,p'}(q)=\sum_s Z_s^{(p,p')}(q)
\end{align}
where $Z_s(q)=\left|\langle s |\rho_q^{(p,p')}|0\rangle\right|^2$. In order to access the properties of supersolidity we need to compute the total static structure factor $S_{tot}(q)= S(q) + S_e(q)$, which takes into account both elastic and inelastic scattering. Inelastic scattering is captured by $S(q)$ and elastic scattering corresponds to the so-called disconnected dynamic structure factor $S_e(q)=Z_{s=0}(q)$~\cite{DynaElastic}.

\section{Excitation spectrum as a probe of the phases of the two-leg bosonic ring ladder }
For the lattice ring three phases are known: the Meissner (M), vortex (V) and biased-ladder (BL) phase. A schematic phase diagram for the infinite-ladder limit   is illustrated in Fig.\ref{PhaseInf}. It is obtained by   minimizing the mean-field energy with respect to the  Ansatz  $|\Psi\rangle = \frac{1}{\sqrt{N!}}\left(\cos(\theta)\hat{\beta}_{k_1}^\dagger+\sin(\theta)\hat{\beta}_{k_2}^\dagger\right)^N|0\rangle$  \cite{Mueller}. The Meissner phase is characterized by vanishing transverse currents $j_{l,\perp}= iK\langle \hat{a}^{\dagger}_{l,1}\hat{a}_{l,2}-\hat{a}^{\dagger}_{l,2}\hat{a}_{l,1}\rangle$; the longitudinal currents on each ring, defined as $j_{l,p}= iJ\langle \hat{a}_{l,p}^{\dagger}\hat{a}_{l+1,p}e^{i\Phi_p}-\hat{a}_{l+1,p}^{\dagger}\hat{a}_{l,p}e^{-i\Phi_p}\rangle$, are opposite and the chiral current, i.e $J_c=\sum_l \langle j_{l,1}-j_{l,2}\rangle$ is saturated. The vortex phase is characterized by a modulated density, jumps of the phase of the wave function, and  non-zero, oscillating  transverse currents which create a  vortex pattern. The biased-ladder phase has only longitudinal currents as in the  Meissner phase, but  displays and imbalanced population between the two rings.

\begin{figure}[htb]
\includegraphics[scale=0.35]{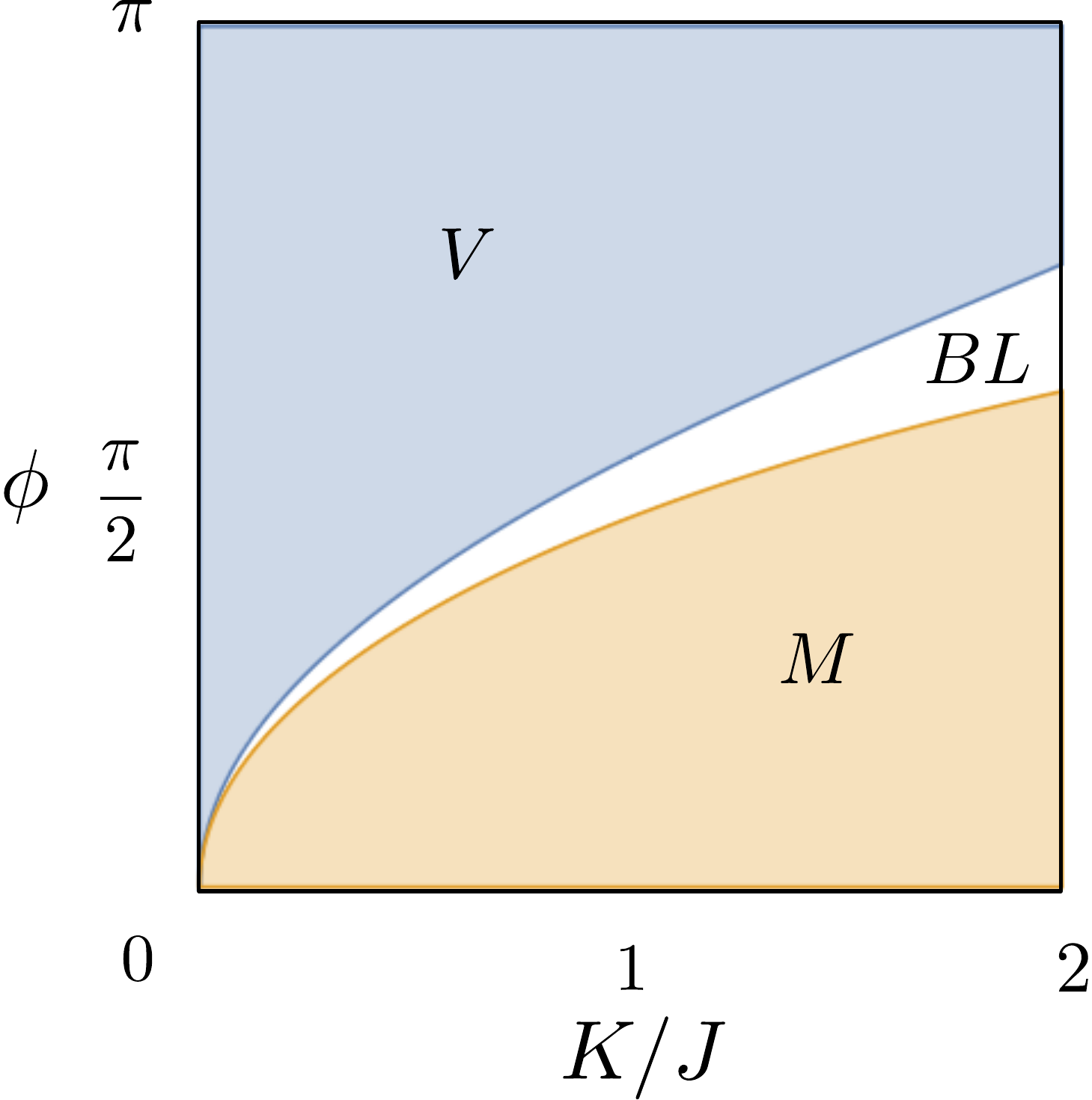}
\label{FigDI}
\caption{Phase diagram of the infinite ladder as a function of the flux $\phi$ per plaquette and tunnel ratio $K/J$ for interaction strength $Un/J=0.01$. The Meissner (M), vortex (V) and biased ladder (BL)  phases are indicated on the figure.}
\label{PhaseInf}
\end{figure}

\subsection{Meissner phase}
In the Meissner phase, the ground-state solution for the condensate wave function is uniform in space and corresponds to a condensate occupying the $k=0$ state.

The lowest branch of the spectrum  shows a single phononic  mode close to $k=0$.
This is the Goldstone mode associated to the spontaneous breaking of the $U(1)$ symmetry. Notice that even though we solve  two equations for the condensate wavefunctions on each ring  (see Eqs.(\ref{dnlse}) and (\ref{BogDe})), only the global phase is free to fluctuate while the relative phase is fixed by the tunnel coupling among the two rings.
A simplified  expression for the dispersion relation  of this branch can be obtained by performing the Bogoliubov approximation on the lower branch of the single-particle spectrum \cite{Mueller}. It reads 
\begin{align}
\epsilon^M_k=\frac{1}{2}\sqrt{(Un+2\tilde{\epsilon}_k)^2-(2Un u_k v_k)^2 }
\label{EqM}
\end{align}
with $\tilde{\epsilon}_k=E_-(k)-E_-(0)$ and $u_k$, $v_k$ defined in Appendix \ref{AppendixA}. An  analytical solution of the full two-band problem is provided in Appendix~\ref{MeissnerExcitation}.

Fig.~\ref{MeisExcFig} shows the dynamical structure factor as  obtained by the numerical diagonalization of the Bogoliubov DeGennes equations. The poles of the dynamical structure factor in the frequency-wavevector plane are in excellent agreement  with Eq.(\ref{EqM}), also shown in the figure.

\begin{figure}[h!]
\includegraphics[scale=1]{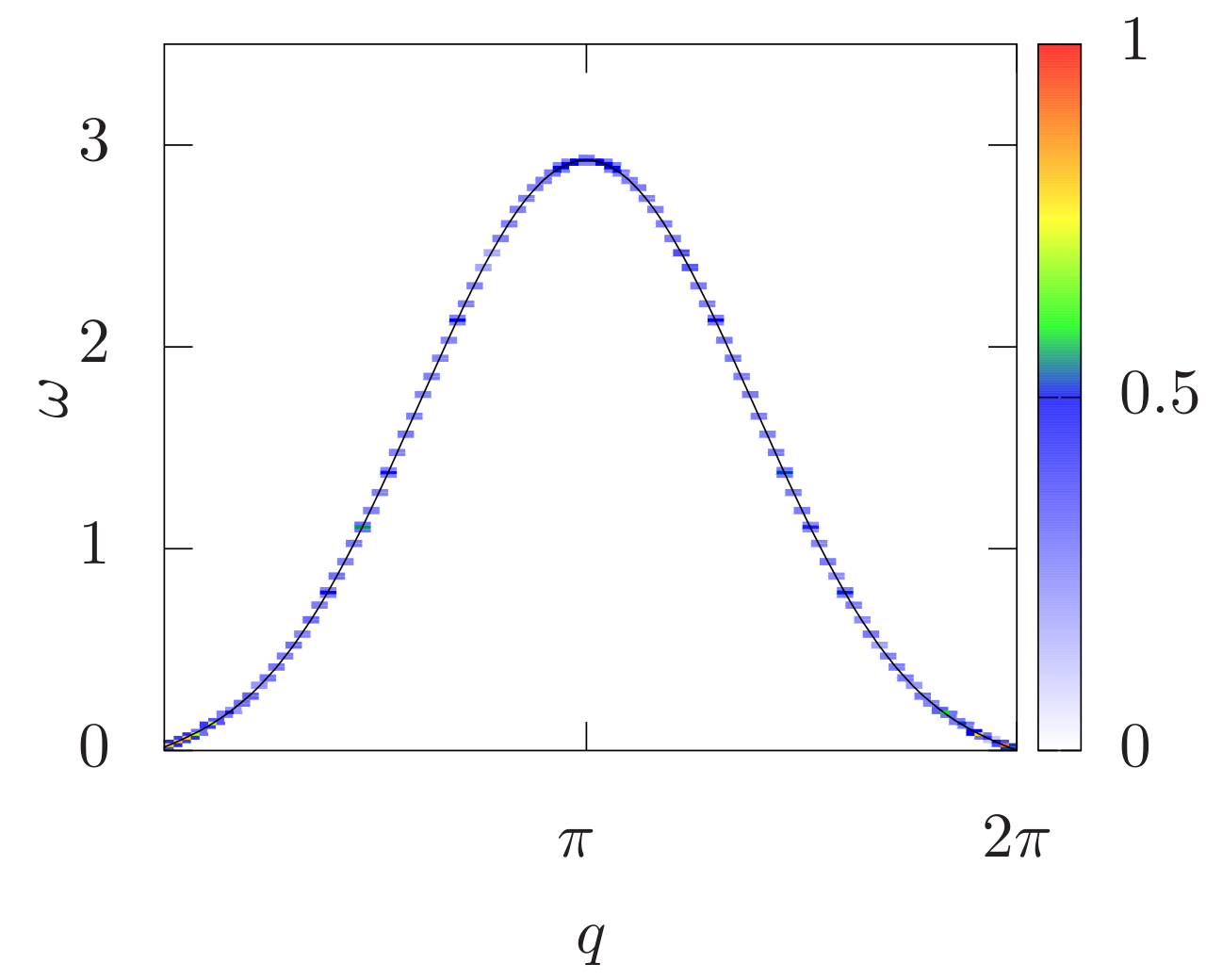}
\caption{Excitation spectrum (solid black line) and dynamical structure factor $S_{\beta\beta}$ in the lowest branch basis in the Meissner phase, in the frequency-wavevector plane (color map, $q$ in units of $1/a$ with $a$ lattice spacing and $\omega$ in units of $J$)  for $Un/J=0.2$, $\phi=\pi/2$, $K/J=3$.}
\label{MeisExcFig}
\end{figure}

\begin{figure}[h!]
\centering
\subfigimg[width=0.26\textwidth]{}{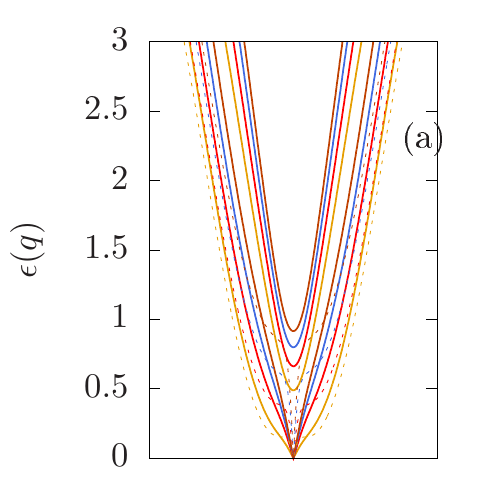}\hspace{-0.8cm}
\subfigimg[width=0.26\textwidth]{}{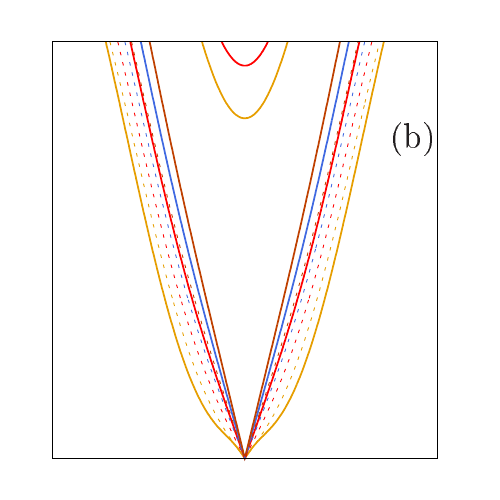}\vspace{-0.5cm}
\subfigimg[width=0.26\textwidth]{}{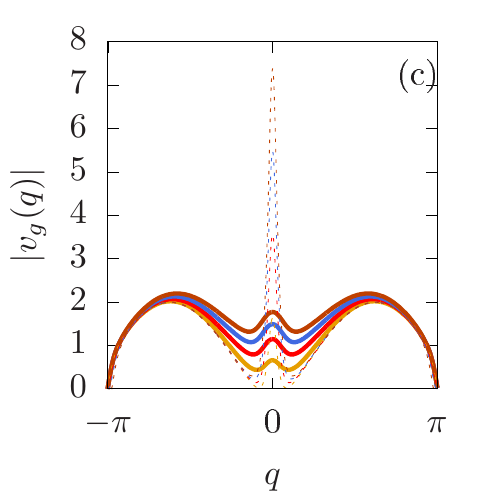}\hspace{-0.8cm}
\subfigimg[width=0.26\textwidth]{}{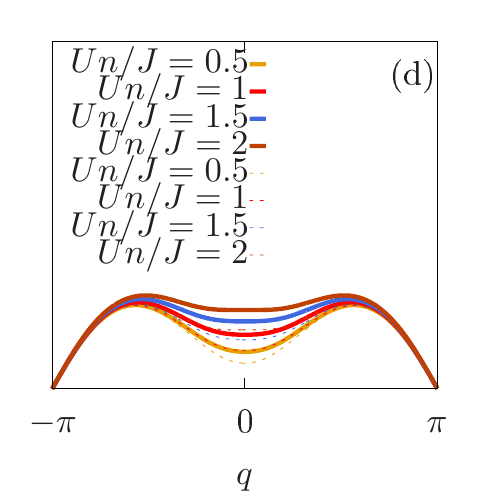}
\label{MeissnerExc}
\caption{Upper panels: dispersion relations in the frequency-wavevector plane  in the Meissner phase for (a) $\phi=\pi/4$, $K/J=0.1$ and (b) $\phi=\pi/4$, $K/J=1$ for various values of the interaction strength indicated on  panel (d). Bottom panels: corresponding group velocity as a function of wave-vector $q$ for (c)  $\phi=\pi/4$, $K/J=0.1$ and (d) $\phi=\pi/4$, $K/J=1$. The solid lines correspond to the Meissner excitation spectrum taking into account both lower and upper branch of the non-interacting problem (see Eq.~\ref{MeissnerExci}), the dashed lines correspond to the lowest band approximation (see Eq.~\ref{EqM}).}
\end{figure}


The dispersion relation $\epsilon_k$ obtained from the full solution  (Eq.~\ref{MeissnerExci}) and the corresponding group velocity  $v_g=\partial \epsilon_k/\partial k$ is shown in Fig.~\ref{MeisExcFig} for various values of the interaction strength. The main effect of the coupling at strong interactions is to change the sound velocity, to decrease the region where the spectrum is linear and modify the shape of the dispersion at finite momenta, where the group velocity displays a minimum.



\subsection{Biased-ladder phase}
In the biased-ladder phase, as well as in the vortex phase,  the single-particle dispersion relation has two minima at  $ k=k_1,k_2$. In the biased-ladder phases only one of the two minima is macroscopically populated.  The excitation spectrum shows a phononic Goldstone mode and a rotonic structure~\cite{Mueller}. A similar behaviour is found in spin-orbit coupled Bose gases \cite{MartonePRA}. At fixed flux $\phi$, when decreasing the coupling $K$ between the ring, the vortex phase is accessed through a softening of the roton minimum. The system enters the vortex phase when the roton minimum decreases down to  a critical (non zero) value thus indicating a first-order transition, similarly to what predicted for dipolar gases \cite{March-Tosi-Anna}.

\begin{figure}[h!]
\centering
\subfigimg[width=0.24\textwidth]{}{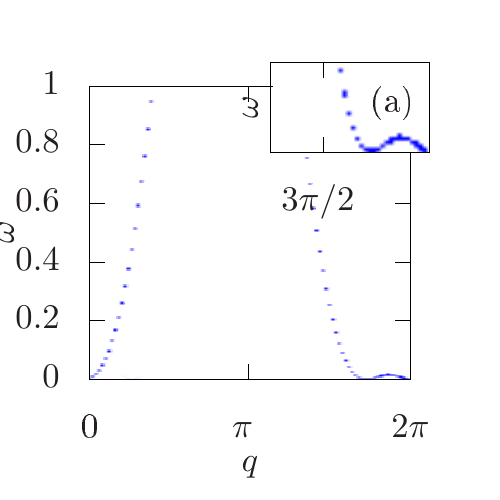}\hspace{-1.4cm}
\subfigimg[width=0.24\textwidth]{}{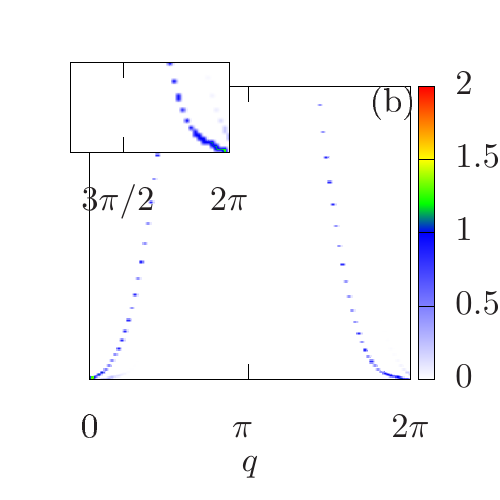}
\caption{(a) Dynamical structure factor for the biased-ladder phase, with $N_s=100$,$Un/J=0.01$,$\phi=\pi/2$ and $K/J=1.15$, (b) dynamical structure factor for the biased-ladder phase, with $N_s=100$,$Un/J=0.05$,$\phi=\pi/2$ and $K/J=1.4$}
\end{figure}

\subsection{Vortex Phase}
At the mean field level, the vortex phase is characterised by a macroscopic occupation  of a superposition state involving two single-particle momentum  modes  $k=k_1,k_2$ corresponding to the minima of the single-particle dispersion relation \cite{Mueller}.

The numerical result for  the dynamical structure factor $S_{\beta,\beta}(q,\omega)$ in the vortex phase  is shown in Fig.\ref{spectreVortex}.  We find various minima of the dispersion relation  for  $q=0$ and $q=k_2-k_1$ as well as at the points $q=2\pi$ and $q=2\pi-(k_2-k_1)$. Two linear dispersion branches are found around each of these  minima, caracterized by two diffferent sound velocities. These branches correspond  to the two Goldstone modes associated to the breaking of U(1) and spatial translational symmetry, and hence may be viewed as phase modes and crystal modes \cite{Saccani2012,Macri2013,Roccuzzo2019}. 

An analysis of the dynamical structure factor in log scale (see Fig. \ref{spectreVortex}.b) shows  a folding of the Brillouin zone for the excitations, corresponding to the underlying ground-state vortex superlattice felt by the Bogoliubov excitations. Specifically, for the parameters chosen in the calculation of Fig.\ref{spectreVortex} we have that the ground state density profile has a modulation with wavevector $k_1-k_2$, leading to a $\frac{2\pi}{k_1-k_2}$-times folding of the excitation spectrum, i.e 5-times in the case of Fig~\ref{spectreVortex}.

The overall features  of the excitation spectrum can be understood by comparing it with  the one in the $K=0$ case. In this regime  the rings are independent and  the excitation spectrum is given by two branches, obtained by solving the Bogoliubov equations for each ring separately:
\begin{widetext}
\begin{align}
\epsilon_k^{V(1)} = -\frac{1}{2}\left(\epsilon_{k+k_1} - \epsilon_{k-k_1} \pm 
   \sqrt{(\epsilon_{k+k_1} + \epsilon_{k-k_1})^2 + 4 U |\psi_0|^2 (\epsilon_{k+k_1} + \epsilon_{k-k_1}})\right)\nonumber\\
\epsilon_k^{V(2)} = -\frac{1}{2}\left(\epsilon_{k+k_2} - \epsilon_{k-k_2} \pm 
   \sqrt{(\epsilon_{k+k_2} + \epsilon_{k-k_2})^2 + 4 U |\psi_0|^2 (\epsilon_{k+k_2} + \epsilon_{k-k_2}})\right)\nonumber\\
\end{align}
\end{widetext}
where  $\epsilon_k=2J(1- \cos(k))$.

Exploiting a  low-energy model (see Appendix \ref{BogoVortexAna})  we can then understand qualitatively the behaviour of the excitation spectrum at small but finite $K$. In this regime, the  excitations can  tunnel from one ring to the other with $k$-dependent interaction parameters $\tilde{U}$ and $\tilde{\tilde{U}}$ (see Appendix \ref{BogoVortexAna}). These scattering events break the degeneracy of the  sound velocities around each minimum.

\begin{figure}[h!]
  \centerline{\includegraphics[scale=0.85]{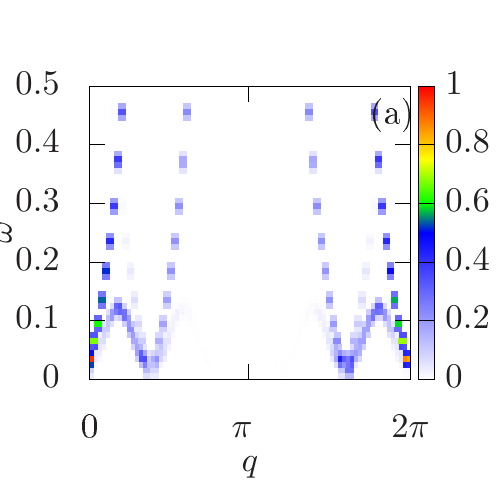}
    \includegraphics[scale=0.85]{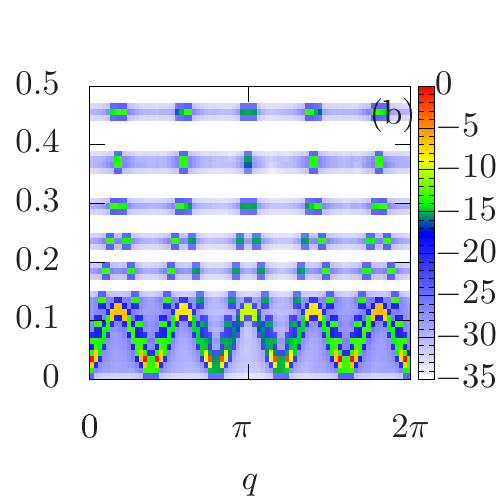}}
    \includegraphics[scale=0.85]{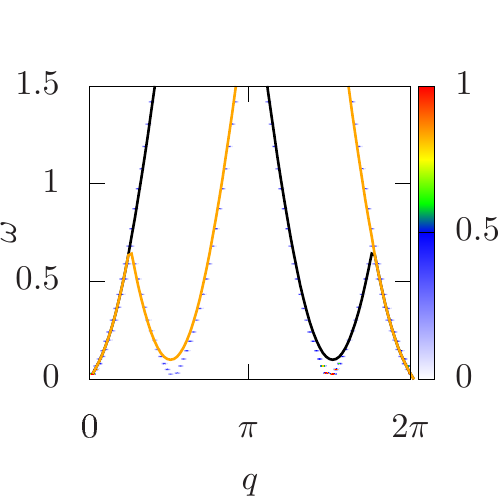}
    \caption{(a) Dynamical structure factor $S_{\beta\beta}(q,\omega)$ in the vortex phase for $K/J=0.8$ in linear (a),  and logarigthmic (b) scale; (c) dynamical structure factor  $S_{\beta\beta}(q,\omega)$  for $K/J = 0$. The other parameters for  all the panels are  $\phi=\pi/2$, $Un/J=0.2$,  $N_s=80$.}
\label{spectreVortex}
\end{figure}

\begin{figure}[h!]
\centering
\subfigimg[width=0.24\textwidth]{}{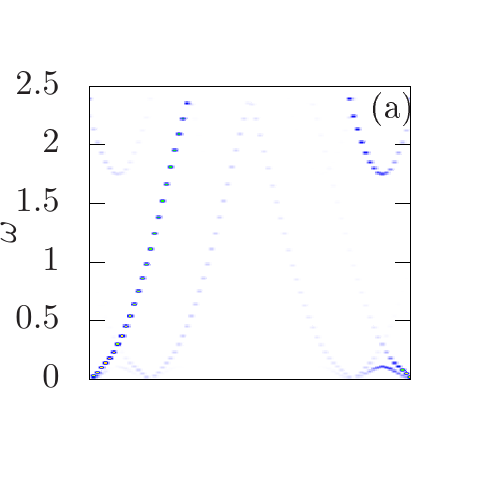}\hspace{-1.4cm}\vspace{-1.5cm}
\subfigimg[width=0.24\textwidth]{}{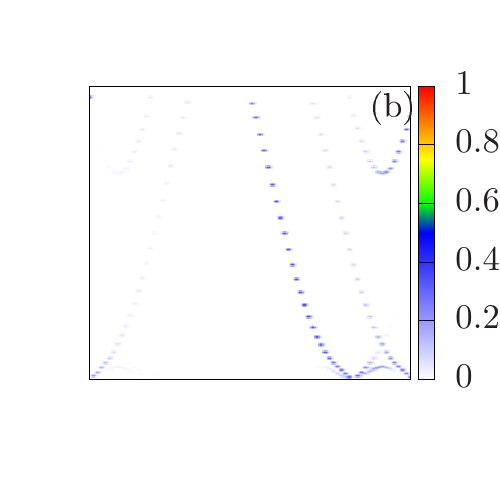}\hspace{-1.4cm}
\subfigimg[width=0.24\textwidth]{}{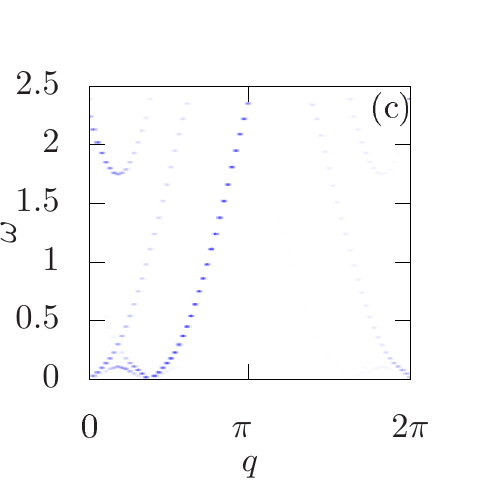}\hspace{-1.4cm}
\subfigimg[width=0.24\textwidth]{}{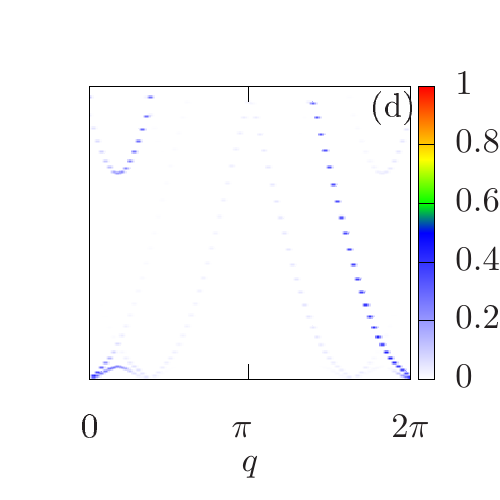}
\caption{Dynamical structure factor $S_{p,p'}(q,\omega)$ in the ring basis   with $p,p'=1,2$ in the vortex phase:  a) $S_{1,1}(q,\omega)$, b) $S_{1,2}(q,\omega)$, c) $S_{2,1}(q,\omega)$ and d) $S_{2,2}(q,\omega)$. For all panels we have taken  $\phi=\pi/2$, $N_s=80$, $K/J=0.8$ and $Un/J=0.2$.}
\label{RingBase}
\end{figure}
In Fig.~\ref{RingBase} we show the various dynamical structure factors $S_{p,p'}(q,\omega)$ in the ring basis. All the excitations branches observed in $S_{\beta\beta}(q,\omega)$ are visible, with variable spectral weight depending on the choice of $p,p'$.
The off-diagonal dynamical structure factors show the symmetry relation $S_{12}(q,\omega)=S_{21}(q,-\omega)$.

\subsection{Experimental probe of the dynamical structure factor}
In ultracold atomic gases  the dynamical structure factor can be measured using two-photon optical Bragg spectroscopy \cite{Brunello}, according to the following scheme: two laser beams are impinged upon the condensate, and the difference in the wave vectors of the beams defines the momentum transfer $\hbar q$, and the frequency  difference defines the energy transfer $\hbar \omega$ to the fluid. Both the values of $q$ and $\omega$ can be tuned by changing the angle between the two beams and varying the frequency difference of the two laser beams. Several experiments have reported the observation of the dynamical structure factor with ultracold atoms (see eg Refs.\cite{KetterleDyna,Landig2015,PhysRevA.91.043617,Natale2019}).

A way of probing the excitation spectrum  of the double ring studied in this work is to use angular momentum spectroscopy  \cite{ProbeDyna}: in this case, one needs  two laser beams  denoted by $1$,$2$ in high-order Laguerre-Gauss modes with optical angular momenta $l_{1,2}$ and frequencies $\omega_{1,2}$. Their  corresponding electric fields read $E_{1,2}(r)=f_{l_1,l_2}(r)e^{-il_{1,2}\theta -i\omega_{1,2}t}$ where the radial mode functions  $f_{l}(r)\propto (r/r_0)^{|l|}e^{r^2/2r^2_0}$  need to be chosen in order to match the shape of the double ring to probe.

\section{Coherence properties and supersolidity}
\subsection{One-body density matrix}
In order to study the coherence properties of the system we consider the one-body density matrix $\rho^{(1)}_{\alpha,\alpha'}(j,l)=\langle \hat{a}_{j,\alpha}^{\dagger}\hat{a}_{l,\alpha'}\rangle$, which in the Bogoliubov approximation reads ~\cite{CastinDum,Ph}
\begin{align}
\rho^{(1)}_{p,p'}(j,l)=\sqrt{\rho^{(0)}_{j,p}\rho^{(0)}_{l,p'}}\exp\left(-\frac{1}{2}\sum_s\left|\frac{Q_{s,j}^{(p)}}{|\Psi^{(0)}_{j,p}|}-\frac{Q_{s,l}^{(p')}}{|\Psi^{(0)}_{l,p'}|}\right|^2\right),
\end{align}
where the function in the exponential relates to the fluctuation of the phase of the condensate and $\rho^{(0)}_{l,p}$ stands for the mean-field ground-state density profile of each ring. Since in the vortex phase the system is inhomogeneous the one-body density matrix does not depend only on the coordinate difference $j-l$. Therefore, to estimate the coherence we study the averaged first-order correlation function defined as 
\begin{align}
g^{(1)}_{(p,p')}(l)=\sum_j\rho^{(1)}_{(p,p')}(j,j+l)/\sqrt{\rho_{j,p}\rho_{j+l,p'}}
\end{align}
\begin{figure}[h!]
\includegraphics[scale=0.9]{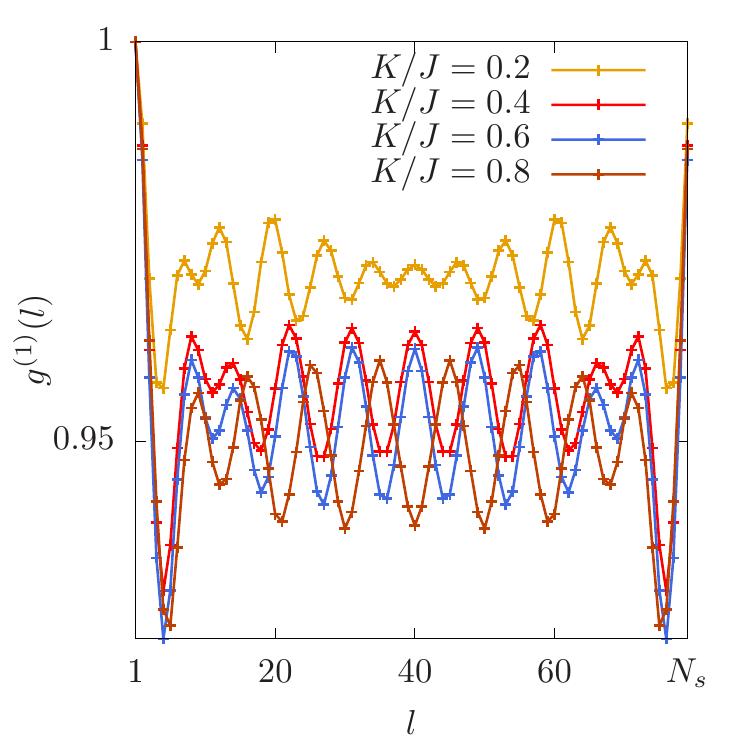}
\caption{Averaged first-order correlation function $g^{(1)}_{(1,1)}(l)$ for the inner ring in the vortex phase for $\phi=\pi/2$, $N_s=80$ and $Un/J=0.2$.}
\label{DegreeV}
\end{figure}
Figure \ref{DegreeV} shows the $g^{(1)}$ correlations in the vortex phase along the inner ring.  As the coupling  $K/J$ between the rings increases,    we notice that the correlations in the ring decrease.
However, even for large values of $K/J$ the coherence in the vortex phase stays high even at large distances. This corresponds to a large  condensate fraction, thereby  implying Bose-Einstein condensation (BEC) and superfluidity.

\subsection{Static structure factor - probe of spatial order}
In order to probe the spatial crystalline order expected in the vortex phase we compute the total static structure factor $S_{tot}(k)$ (see Sec.IIC), which is illustrated in  Fig.\ref{Stot}. We clearly see a peak at wavevectors $k=k_2-k_1$ and $k=2\pi-(k_2-k_1)$, revealing the crystalline order associated to the spatial  modulations of the condensate density profile.

\begin{figure}[h!]
\centering
\includegraphics[scale=0.6]{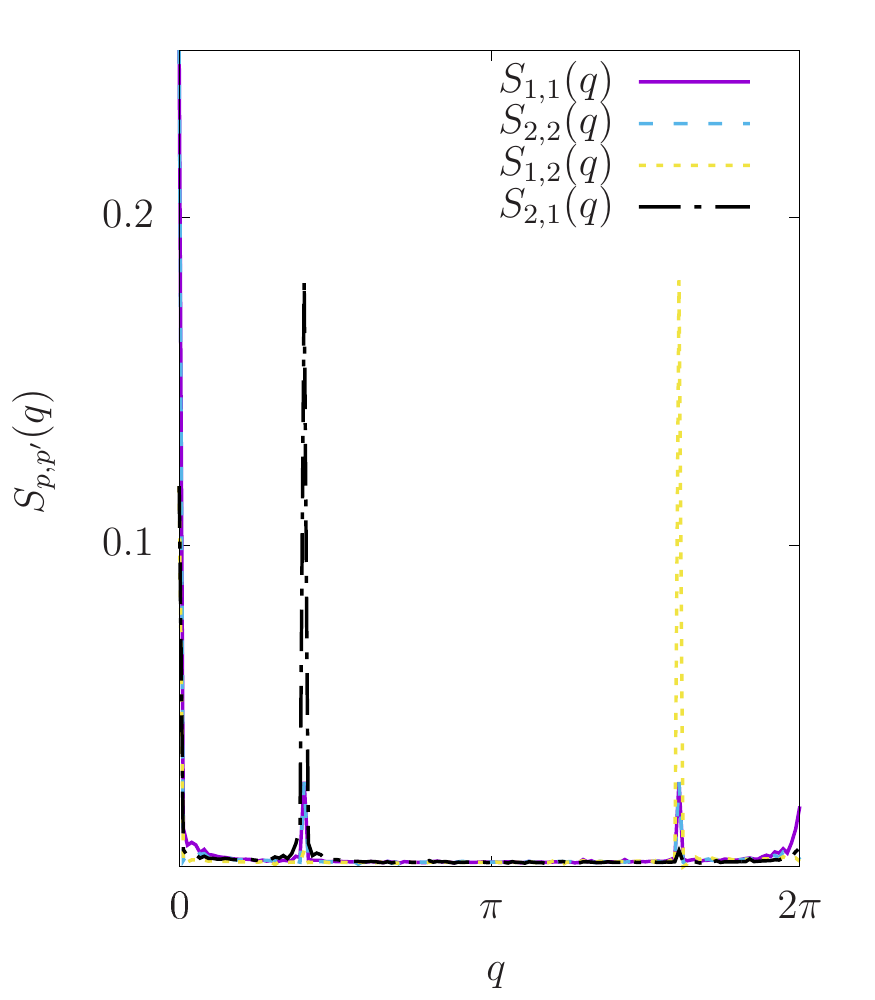}
\label{Stot}
\caption{Total static structure factor $S_{tot}(q)$ as a function of wavevector $q$ (in units of $1/a$) for $\phi=\pi/2$, $N_s=150$, $K/J=0.8$ and $Un/J=0.2$.}
\end{figure}

\section{Small ring limit and nature of the excitations}
We report in this section a study on the nature of the excitations in the different phases of the system. 

For this purpose, we calculate the density fluctuations  $\delta n_{l,p}^{\nu}$ defined as $\langle s |\rho_{l,p}|0\rangle$ which can be obtained from  of the Bogoliubov eigenmodes $h_{\nu,l}^{(p)}$ and $Q_{\nu,l}^{(p)}$ according to 
 \begin{equation}
 \delta n_{l,p}^{\nu}=2\, {\mathrm Re}\left[\Psi_{l,p}^{(0)}(h_{\nu,l}^{(p)})^*-(\Psi_{l,p}^{(0)})^*Q_{\nu,l}^{(p)}\right].
 \end{equation}

 Our results for the density fluctuations of chosen low-energy modes are shown in Fig.~\ref{VortexJ}.  Among the various types of excitation modes, in addition to the phononic Goldstone modes propagating along each ring, we identify the Josephson mode, typical of a finite ring system, which is characterized by spatially homogeneous density fluctuations and out-phase oscillations of the relative populations among  the two rings, as in the small-amplitude dynamics of the Josephson effect \cite{JosephsonBogoModugno,Paraoanu_2001}.
 
 We see  in Fig~\ref{VortexJ} that a uniform Josephson mode occurs at low energy in the Meissner phase for low enough coupling among the rings,  whereas higher excited mode are of phonons of  charge (ie in-phase)  type. Close to  the phase boundary, in the vortex phase we find that the lowest excitation is a spin (ie out-of-phase)  oscillation.  In the nearby Meissner phase the lowest excitation become phonon of charge type, as well as in the biased-ladder phase.

\begin{figure}
\includegraphics[scale=0.70]{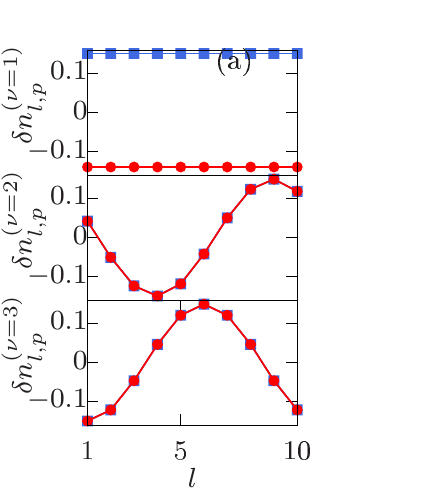}\hspace{-1.42cm}
\includegraphics[scale=0.70]{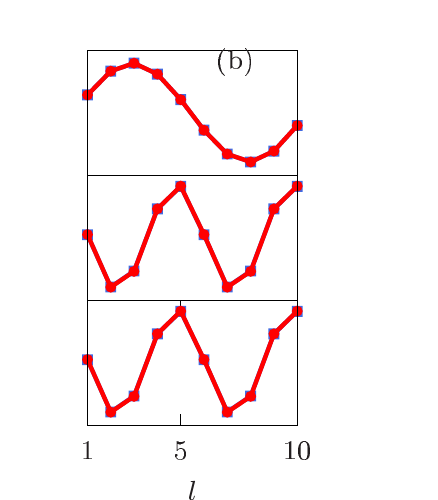}\hspace{-1.42cm}
\includegraphics[scale=0.70]{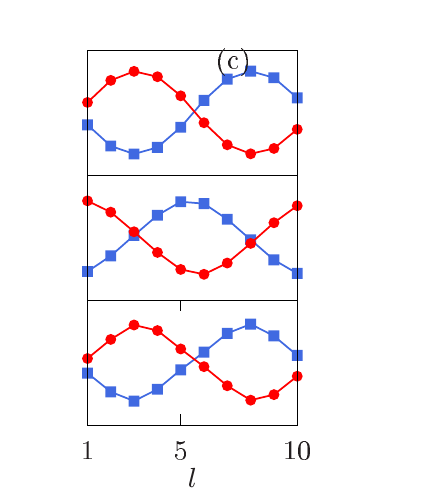}\hspace{-1.42cm}
\includegraphics[scale=0.70]{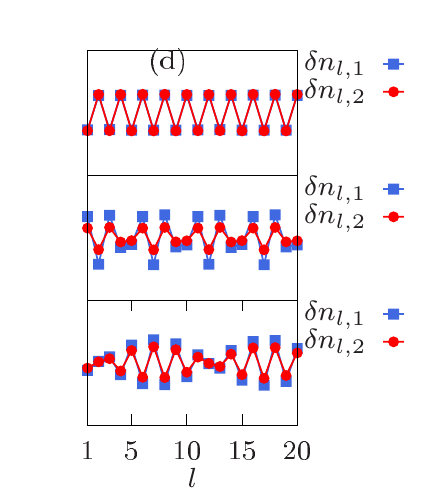}
\caption{Excitation eigenmodes $\delta n_{l,p}$ as a function of the position $l$ along each ring (blue lines with squares $p=1$, red lines with squares $p=2$)  for the first three excited Bogoliubov modes.   a) Josephson out-of-phase  mode and two in-phase phonon modes  in  the Meissner phase  for $N_s=10$, $K/J=0.1$, $\phi=0.1$, $Un/J=0.1$; b) in-phase phonon modes  in the Meissner phase for $N_s=10$ , $K/J=0.8$, $\phi=0.1$, $Un/J=0.2$; c) out-of-phase   mode  in the vortex phase for  $N_s=10$, $K/J=0.01$, $\phi=\pi/4$ ,$Un/J=0.2$; d) in-phase modes in the biased-ladder phase with $N_s=20$, $K/J=1$ and $\phi=\pi/2$.}
\label{VortexJ}
\end{figure}

The  Josephson modes are found in the Meissner phase for weak tunnel coupling $K/J$ and weak flux $\phi$.  In order to estimate the parameter regime where phonon or Josephson modes are present in the ring, we provide here below some estimates based on energy scales.
In the Meissner phase, close to $k \rightarrow 0$ 
the spectrum has a linear behaviour, 
\begin{align}
\epsilon_{k}^{M}\approx E_{\mathrm{ph}}k 
\end{align}
where
\begin{eqnarray}
 E_{\mathrm{ph}}\!\!=\! \frac{2\pi J^2}{K N_s}\sqrt{\frac{U}{J}\left[\frac{K^2}{J^2} \cos(\phi/2) + \left(\frac{U}{J}-\frac{2K}{J}\right) \sin(\phi/2)^2\right]}.
 \end{eqnarray}
When comparing it to the energy of the Josephson mode which scales as the band gap between the upper and lower branch of the excitation spectrum $E_{\mathrm{gap}} \approx K$ we predict that the region where Josephson modes are allowed, ie when $ E_{\mathrm{gap}} < E_{\mathrm{ph}}$,   appears at very low $K$ and $\phi$ (see Fig.~\ref{Scaling}). This is in agreement with the numerical simulations. Moreover, we obtain that Josephson region shrinks at increasing the number of sites in the ring, thereby showing that the Josephson modes are a finite-size effect.

\begin{figure}
\centering
\includegraphics[scale=0.30]{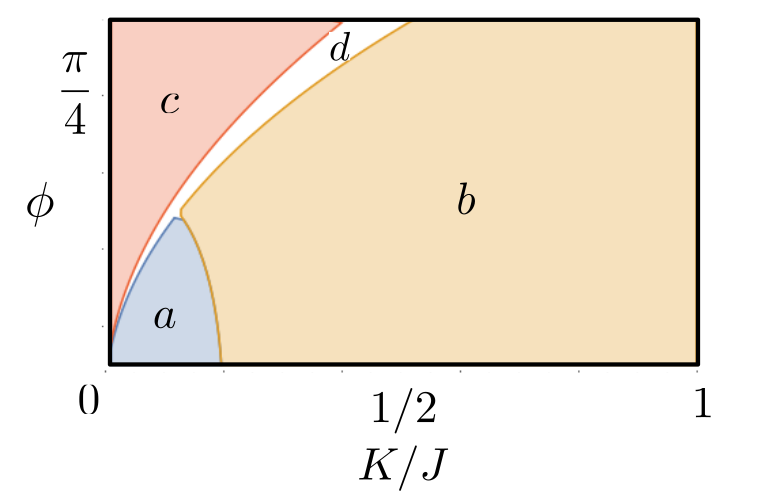}
\caption{Sketch of phase diagram at fixed interaction strengh $Un/J=0.2$ and $N_s=10$ as deduced from the analysis of the excitation eigenmodes, summarizing the cases illustrated in Fig.~\ref{VortexJ}. The coloured regions marked by letters indicate: a) Meissner phase with lowest mode of Josephson type, b) Meissner phase with lowest mode of charge type, c) Vortex phase with lowest mode of spin type, d) Biased-ladder phase with lowest mode of charge type.}
\label{Scaling}
\end{figure}

\section{Conclusions}

In conclusion, in this work we have performed  a detailed study of the excitation spectrum of a weakly interacting Bose gas in a two-leg bosonic ring ladder subjected to two artificial gauge fields. For all the three phases expected at weak interactions, i.e the Meissner, vortex and biased-ladder phase, we have solved  the Bogoliubov-de Gennes equations for the ring ladder and  calculated the dynamical structure factor. For a cigar-shaped gas and for a one-dimensional gas in a linear atomic waveguide,  the dynamical structure factor has been already experimentally measured. Here we propose that it is accessed in the ring geometry by angular momentum spectroscopy.

Our main predictions are a single phonon-like dispersion at long wavelength in the Meissner phase, a roton minimum emerging in the biased-ladder phase and two  phononic branches in the vortex phase. Furthermore, we find evidence of the underlying spatially modulated structure of the vortex phase in the spectrum by a folding of the Brillouin zone of the excitations.

Using the Bogoliubov excitations eigenmodes, we have also calculated the first-order correlation function, monitoring the coherence of the gas, and found that it remains high all over the rings.  This feature,  together with the diagonal long-range order in the vortex phase is hallmarking the supersolid nature of the fluid. The emergence of supersolidity in this system is quite remarkable, as, at difference from the spin-orbit coupled Bose gas,  the visibility of the fringes can be  tuned thanks to the absence of interspecies contact interactions in the current model. Finally, we have shown the emergence of Josephson  excitations in a finite ring, corresponding to population imbalance oscillations among the two rings

In outlook, it would be interesting to study the excitation spectrum at larger interaction strengths, where the nature of the ground state changes onto a fragmented condensate \cite{Kolovsky2017} or a fragmented Fermi sphere \cite{Victorin2019} at intermediate and large interactions respectively. The knowledge of the excitation spectrum is also useful for atomtronics appications \cite{Amico_Atomtronics,Amico_focus}, eg for the study of transport in the linear-response regime,  when two leads are attached to the ring \cite{PhysRevLett.74.1843,PhysRevLett.100.140402,Haug_2019}.

\acknowledgements
We thank L. Amico and S. Moroni for stimulating discussions. We acknowledge funding from the ANR SuperRing (Grant No.  ANR-15-CE30-0012).

\bibliographystyle{prsty}
\bibliography{BibTexNotes,BibBogoadd}

\begin{thebibliography}{10}

\bibitem{LegettSuper}
A.~J. Leggett, Phys. Rev. Lett. {\bf 25},  1543  (1970).

\bibitem{Chester}
G.~V. Chester, Phys. Rev. A {\bf 2},  256  (1970).

\bibitem{DickeQuantumPhase}
B. Kristian, G. Christine, B. Ferdinand, and E. Tilman, Nature {\bf 464},
  1301–1306  (2010).

\bibitem{Leonard2017a}
J. L{\'e}onard, A. Morales, P. Zupancic, and T.~E. andT. Donner, Nature {\bf
  543},  87  (2017).

\bibitem{Leonard2017b}
J. L{\'e}onard {\it et~al.}, Science {\bf 543},  91  (2017).

\bibitem{Li2017}
J.-R. Li {\it et~al.}, Nature {\bf 543},  91  (2017).

\bibitem{Boettcher2019}
F. B{\"o}ttcher {\it et~al.}, Phys Rev X {\bf 9},  011051  (2019).

\bibitem{Tanzi2019}
L. Tanzi {\it et~al.}, Phys. Rev. Lett. {\bf 122},  130405  (2019).

\bibitem{Chomaz2019}
L. Chomaz {\it et~al.}, Phys. Rev. X {\bf 9},  021012  (2019).

\bibitem{Natale2019}
G. Natale {\it et~al.}, Excitation Spectrum of a Trapped Dipolar Supersolid and
  Its Experimental Evidence, 2019.

\bibitem{Quantumphasesfromcompetingshort-andlong-rangeinteractions}
L. Renate {\it et~al.}, Nature {\bf 532},  476–479  (2016).

\bibitem{Gross}
E.~P. Gross, Phys. Rev. {\bf 106},  161  (1957).

\bibitem{Spinorbit}
Y. Li, G.~I. Martone, L.~P. Pitaevskii, and S. Stringari, Phys. Rev. Lett. {\bf
  110},  235302  (2013).

\bibitem{SpinoExp}
J.-R. Li {\it et~al.}, Nature {\bf 543},  91 EP   (2017).

\bibitem{Ketterle}
W. Ketterle and N.~J. van Druten, Phys. Rev. A {\bf 54},  656  (1996).

\bibitem{PhysRevB.91.140406DMRG}
M. Piraud {\it et~al.}, Phys. Rev. B {\bf 91},  140406  (2015).

\bibitem{Georges}
A. Tokuno and A. Georges, New Journal of Physics {\bf 16},  073005  (2014).

\bibitem{PhysRevB.64.144515Orignac}
E. Orignac and T. Giamarchi, Phys. Rev. B {\bf 64},  144515  (2001).

\bibitem{ChiralMottDhar}
A. Dhar {\it et~al.}, Phys. Rev. B {\bf 87},  174501  (2013).

\bibitem{PhysRevLett.111.150601}
A. Petrescu and K. Le~Hur, Phys. Rev. Lett. {\bf 111},  150601  (2013).

\bibitem{Mueller}
R. Wei and E.~J. Mueller, Phys. Rev. A {\bf 89},  063617  (2014).

\bibitem{Atala}
M. Atala {\it et~al.}, Nature Physics {\bf 10},  588 EP   (2014), article.

\bibitem{Stuhl}
B.~K. Stuhl {\it et~al.}, Science {\bf 349},  1514  (2015).

\bibitem{Mancini}
M. Mancini {\it et~al.}, Science {\bf 349},  1510  (2015).

\bibitem{Fetter}
A.~L. Fetter, Annals of Physics {\bf 70},  67  (1972).

\bibitem{Dyna}
F. Zambelli, L. Pitaevskii, D.~M. Stamper-Kurn, and S. Stringari, Phys. Rev. A
  {\bf 61},  063608  (2000).

\bibitem{DynaElastic}
F. Zambelli, L. Pitaevskii, D.~M. Stamper-Kurn, and S. Stringari, Phys. Rev. A
  {\bf 61},  063608  (2000).

\bibitem{MartonePRA}
G.~I. Martone, Y. Li, L.~P. Pitaevskii, and S. Stringari, Phys. Rev. A {\bf
  86},  063621  (2012).

\bibitem{March-Tosi-Anna}
A. Minguzzi, N.~H. March, and M.~P. Tosi, Phys. Rev. A {\bf 70},  025601
  (2004).

\bibitem{Saccani2012}
S. Saccani, S. Moroni, and M. Boninsegni, Phys. Rev. Lett. {\bf 108},  175301
  (2012).

\bibitem{Macri2013}
T. Macr\`{\i}, F. Maucher, F. Cinti, and T. Pohl, Phys. Rev. A {\bf 87},
  061602  (2013).

\bibitem{Roccuzzo2019}
S.~M. Roccuzzo and F. Ancilotto, Supersolid behavior of a dipolar Bose-Einstein
  condensate confined in a tube, 2019.

\bibitem{Brunello}
A. Brunello {\it et~al.}, Phys. Rev. A {\bf 64},  063614  (2001).

\bibitem{KetterleDyna}
D.~M. Stamper-Kurn {\it et~al.}, Phys. Rev. Lett. {\bf 83},  2876  (1999).

\bibitem{Landig2015}
R. Landig {\it et~al.}, Nature Communications {\bf 6},  7046  (2015).

\bibitem{PhysRevA.91.043617}
N. Fabbri {\it et~al.}, Phys. Rev. A {\bf 91},  043617  (2015).

\bibitem{ProbeDyna}
N. Goldman, J. Beugnon, and F. Gerbier, Phys. Rev. Lett. {\bf 108},  255303
  (2012).

\bibitem{CastinDum}
Y. Castin and R. Dum, Phys. Rev. Lett. {\bf 77},  5315  (1996).

\bibitem{Ph}
L. Fontanesi, M. Wouters, and V. Savona, Phys. Rev. Lett. {\bf 103},  030403
  (2009).

\bibitem{JosephsonBogoModugno}
A. Burchianti, C. Fort, and M. Modugno, Phys. Rev. A {\bf 95},  023627  (2017).

\bibitem{Paraoanu_2001}
G.-S. Paraoanu, S. Kohler, F. Sols, and A.~J. Leggett, Journal of Physics B:
  Atomic, Molecular and Optical Physics {\bf 34},  4689  (2001).

\bibitem{Kolovsky2017}
A.~R. Kolovsky, Phys. Rev. A {\bf 95},  033622  (2017).

\bibitem{Victorin2019}
N. Victorin {\it et~al.}, Phys. Rev. A {\bf 99},  033616  (2019).

\bibitem{Amico_Atomtronics}
L. Amico and A. M.~G.~Boshier, {\em R. Dumke, Roadmap on quantum optical
  systems} (Journal of Optics, ADDRESS, 2016), Vol.~18, pp.\ 093001,
  doi:10.1088/2040--8978/18/9/093001.

\bibitem{Amico_focus}
L. Amico, G. Birkl, M. Boshier, and L.-C. Kwek, New Journal of Physics {\bf
  19},  020201  (2017).

\bibitem{PhysRevLett.74.1843}
R. Fazio, F.~W.~J. Hekking, and A.~A. Odintsov, Phys. Rev. Lett. {\bf 74},
  1843  (1995).

\bibitem{PhysRevLett.100.140402}
A. Tokuno, M. Oshikawa, and E. Demler, Phys. Rev. Lett. {\bf 100},  140402
  (2008).

\bibitem{Haug_2019}
T. Haug, R. Dumke, L.-C. Kwek, and L. Amico, Quantum Science and Technology
  {\bf 4},  045001  (2019).

\bibitem{Oktel}
A. Kele\ifmmode~\mbox{\c{s}}\else \c{s}\fi{} and M.~O. Oktel, Phys. Rev. A {\bf
  91},  013629  (2015).

\end{thebibliography}

\appendix
\section{Non interacting regime}
\label{AppendixA}
We first proceed by analyzing the non-interacting problem.
The diagonalization of $H_0$ (see Appendix A for details) yields the following two-band Hamiltonian:
\begin{eqnarray}
\hat{H}_0 = \sum_k \hat{\alpha}_k^{\dagger}\hat{\alpha}_kE_+(k) + \hat{\beta}_k^{\dagger}\hat{\beta}_k E_-(k),
\end{eqnarray}
where,
\begin{align}
\begin{pmatrix}
\hat{a}_{k,1}\\
\hat{a}_{k,2}
\end{pmatrix}
=
\begin{pmatrix}
v_k & u_k\\
-u_k & v_k
\end{pmatrix}
\begin{pmatrix}
\hat{\alpha}_k\\
\hat{\beta}_k
\end{pmatrix},
\label{Diag}
\end{align}
where functions $u_k$ and $v_k$ depend on the parameter $\phi$ and $K/J$, and are given here for simplicity in the case $\Phi=0$ trated in this work,
\begin{eqnarray}
v_k = \sqrt{\frac{1}{2}(1+\frac{\sin(\phi/2)\sin(k)}{\sqrt{(K/2J)^2+\!\sin^2(\phi/2)\sin^2(k))}})}\\
u_k = \sqrt{\frac{1}{2}(1-\frac{\sin(\phi/2)\sin(k)}{\sqrt{(K/2J)^2+\!\sin^2(\phi/2)\sin^2(k))}})}.
\end{eqnarray}
the momentum in units of inverse lattice spacing takes discrete values given by  $k =\frac{2 \pi n}{N_s}$, with $n=0,1,2... N_s-1$  and the dispersion relation    $E_{\pm}$ reads 
\begin{align}
E_{\pm}(k)=-&2J\cos(\phi/2)\cos(k)\nonumber\\ \pm &\sqrt{K^2+(2J)^2\sin(\phi/2)^2\sin(k)^2}.
\label{eq:epm}
\end{align}

\section{Excitation spectrum in the Meissner phase}
\label{MeissnerExcitation}
In the Meissner phase, the mean-field solution is uniform in space so that $\Psi_{l,p}=\sqrt{N/2N_s}=\sqrt{n}$. The equation of motion for the Bogoliubov modes expanded in plane wave solutions $h_{\nu,l}^{(p)}=h_{q}^{(p)}e^{iql}$, $Q_{\nu,l}^{(p)}=Q_{q}^{(p)}e^{iql}$,   where $\nu$ is identified as the plane-wave momentum $q$ logitudinal to the rings. It reads 
\begin{widetext}
\begin{align}
\epsilon(q) \begin{pmatrix}
h_{q}^{(1)}\\
Q_{q}^{(1)}\\
h_{q}^{(2)}\\
Q_{q}^{(2)}
\end{pmatrix}
=\begin{pmatrix}
\epsilon_+(q)+Un & -Un & -K & 0 \\
Un & -\epsilon_-(q)-Un &0&K \\
-K &0&\epsilon_-(q) + Un & -Un \\
0& K& Un & -\epsilon_+(q)+Un
\end{pmatrix}
\begin{pmatrix}
h_{q}^{(1)}\\
Q_{q}^{(1)}\\
h_{q}^{(2)}\\
Q_{q}^{(2)}
\end{pmatrix}
\end{align}
\end{widetext}
where $\epsilon_{\pm}(q)= -2J[\cos(k\pm\phi/2)-\cos(\phi/2)]+K$. 

This matrix is diagonalizable and the positive eigenvalues read
\begin{widetext}
\begin{eqnarray}
&\epsilon(q)=\frac{1}{\sqrt{2}}\left\{\epsilon_{+}^2+\epsilon_-^2+2\left(\epsilon_{+}+\epsilon_-\right)Un+2K^2\right.\nonumber\\
&\left.\pm\sqrt{\left(\epsilon_{+}^2-\epsilon_-^2\right)^2+4Un\left(\epsilon_+^3+\epsilon_-^3-\epsilon_-^2\epsilon_++\epsilon_+^2\epsilon_- \right)+4K^2\left[\left(\epsilon_{+}+\epsilon_-\right)^2+4Un\left(\epsilon_{+}+\epsilon_-+Un\right)\right]}\right\}^{1/2}
\label{MeissnerExci}
\end{eqnarray}
\end{widetext}
For a similar derivation see  Ref.\cite{Oktel}.

\section{Bogoliubov excitation spectrum for the lowest single-particle branch}
\label{BogoVortexAna}
In this part, using the following ansatz  (\ref{Ansatz})
\begin{align}
|\Psi\rangle = \frac{1}{\sqrt{N!}}\left( \hat{\beta}_{k_1}^{\dagger}e^{-i\psi_1}+\hat{\beta}_{k_2}^{\dagger}e^{-i\psi_2}\right)^N|0\rangle
\label{Ansatz}
\end{align}
for the ground state, we study the excitation spectrum of the vortex phase by analyzing the Bogoliubov excitations on top of the lowest single-particle excitation branch $\beta$.  The contributions from  the upper  branch, which corresponds to particles created by the operators $\hat{\alpha}_k$, are negligible when the  interaction strength is much smaller than  the  gap among the lower and upper branch of the single-particle spectrum.

In order to perform the Bogoliubov analysis we start from the original Hamiltonian (\ref{eq1}) and compute the interacting part of the Hamiltonian in the free particle diagonal basis $\{\beta_k^{\dagger},\beta_k\}$ (see \cite{Georges}). we obtain
\begin{align}
\hat{H}_{int}=\frac{U}{2N_s}\sum_{q,k,r}K(k-q,r+q,k,r)\hat{\beta}^{\dagger}_{k-q}\hat{\beta}^{\dagger}_{r+q}\hat{\beta}_{k}\hat{\beta}_r
\end{align}
where the kernel $K$ is given by $K(q_1,q_2,q_3,q_4)=u_{q_1}u_{q_2}u_{q_3}u_{q_4}+v_{q_1}v_{q_2}v_{q_3}v_{q_4}$.\\
Here we see that the restriction to the lowest branch yields a one-dimensional Bose gas with effective non-zero range interaction potential. We then proceed by performing the Bogoliubov approximation: we assume that the states $k_1$ and $k_2$ are macroscopically occupied and so approximate the operators in those states by $\mathbb{C}$-numbers:
\begin{align}
\beta_{k_1}=\sqrt{N_0/2}e^{i\psi_1}\\
\beta_{k_2}=\sqrt{N_0/2}e^{i\psi_2}
\end{align}
where $N_0$ is the number of condensed particles in the whole system. We then rewrite the Hamiltonian keeping up all terms up to quadratic order in operators $\hat{\beta}_{k\neq{k_{1},k_{2}}}$, $\hat{\beta}^{\dagger}_{k\neq{k_{1},k_2}}$. In order to conserve particle number within the Bogoliubov approximation we write the number of condensed particles as a function of the total particle number using relation $N_0 =N-\sum_{k\neq (k_1,k_2)}\hat{\beta}^{\dagger}_{k}\hat{\beta}_{k}$. This procedure yields the following quadratic Hamiltonian:
\begin{align}
\hat {H}=E^{(0)}+\hat{H}_{Bog},
\end{align}
where $H^{(0)}$ is the mean field energy in the vortex phase given by:
\begin{align}
&E^{(0)}= NE_{-}(k_1)+\frac{UNn}{4}\left[1+2u_{k_1}^2v_{k_1}^2\right]\\
&\hat{H}_{Bog}=\sum_{k\neq (k_1,k_2)} \tilde{\epsilon}_k\hat{\beta}^{\dagger}_{k}\hat{\beta}_{k}+\sum_k\hat{\beta}_{2k_1+k}^{\dagger}\hat{\beta}^{\dagger}_{-k}U_{1,k}+h.c\nonumber\\&+\sum_k\hat{\beta}^{\dagger}_{2k_2+k}\hat{\beta}^{\dagger}_{-k}U_{2,k}+h.c
+\sum_k\hat{\beta}^{\dagger}_{k_1+k_2+k}\hat{\beta}^{\dagger}_{-k}U_{12,k}+h.c\nonumber\\&+\sum_k\hat{\beta}^{\dagger}_{k_1-k_2+k}\hat{\beta}_{k}(\tilde{U}_{12,k}+c.c)+\sum_k\hat{\beta}_{k_2-k_1+k}^{\dagger}\hat{\beta}_{k}(\tilde{\tilde{U}}_{12,k}+c.c)\nonumber
\end{align}
and the coefficients $\tilde{\epsilon}_k$ and $U_k$  correspond to the Feynman diagrams of Fig.~\ref{Feymann} and read
\begin{align}
&\tilde{\epsilon}_k=E_{-}(k)-E_-(k_1)-\frac{UN}{2N_s}\left(1+2u_{k_1}^2v_{k_1}^2\right)\nonumber\\
&+\frac{UN}{N_s}\left(u_{k_1}^2u_k^2+v_{k_1}^2v_k^2\right)+\frac{UN}{N_s}\left(u_{k_2}^2u_k^2+v_{k_2}^2v_k^2\right)\\
&U_{1,k}=\frac{UN}{4N_s}K(k_1,k_1,2k_1+k,-k)\\
&U_{2,k}=\frac{UN}{4N_s}K(k_2,k_2,2k_2+k,-k)\\
&U_{12,k}=\frac{UN}{4N_s}2K(k_1+k_2+k,-k,k_1,k_2)\\
&\tilde{U}_{12,k}=\frac{2UN}{N_s}K(k_1,k_2,k_1-k_2+k,k)\\
&\tilde{\tilde{U}}_{12,k}=\frac{2UN}{N_s}K(k_1,k_2,k_2-k_1+k,k).
\end{align}

\begin{figure}
\centering
\includegraphics[scale=0.12]{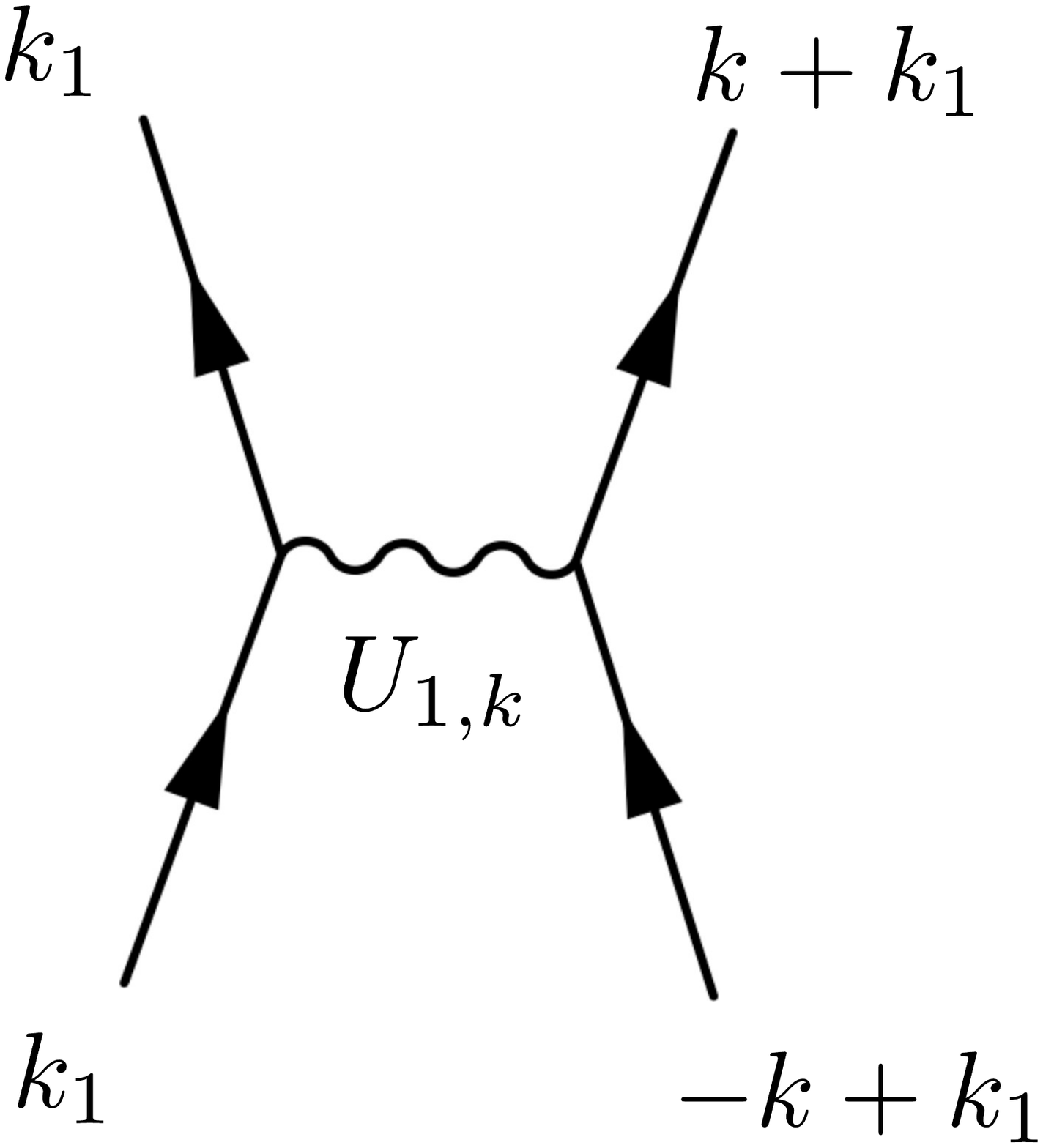}
\includegraphics[scale=0.12]{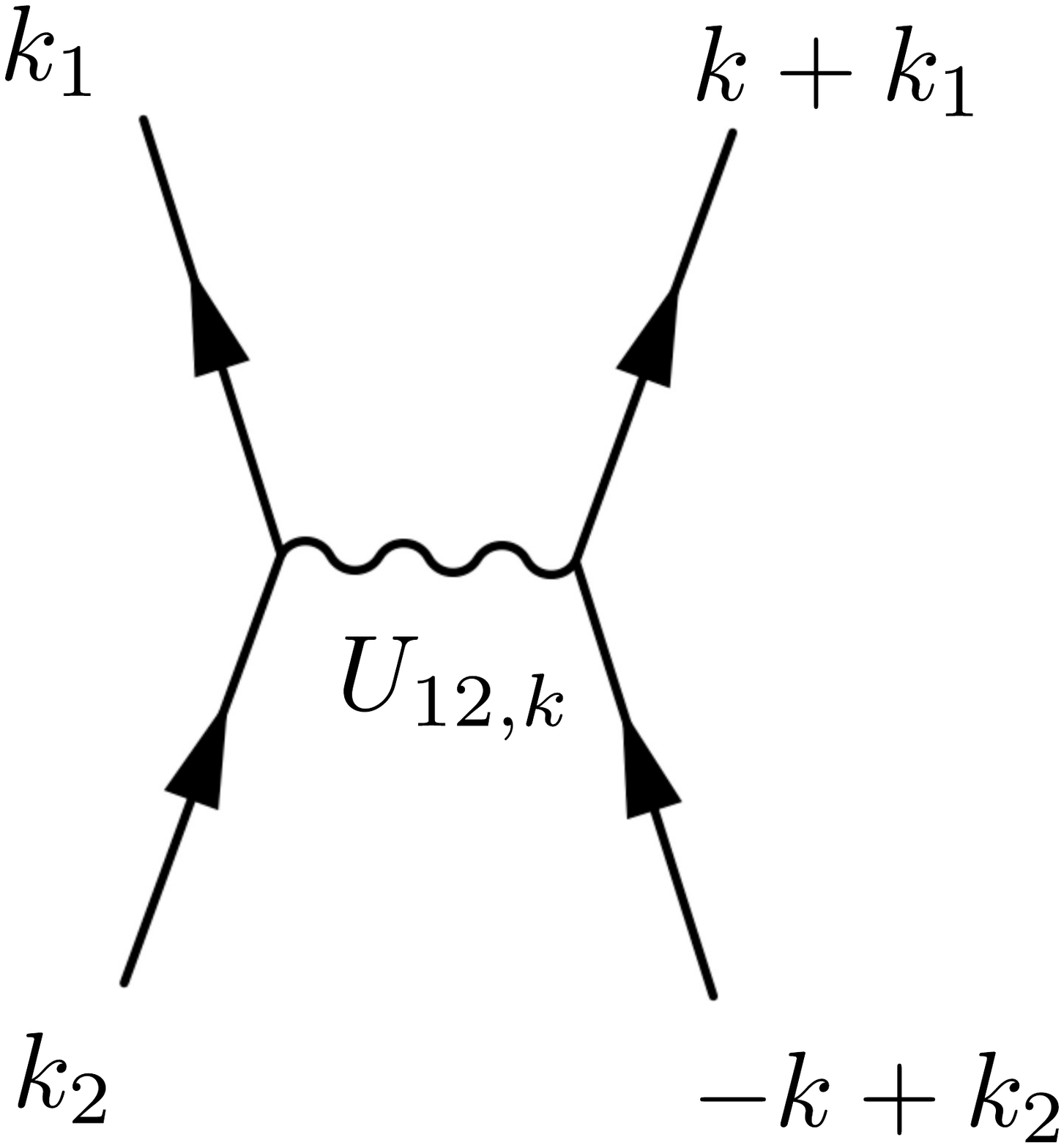}
\includegraphics[scale=0.12]{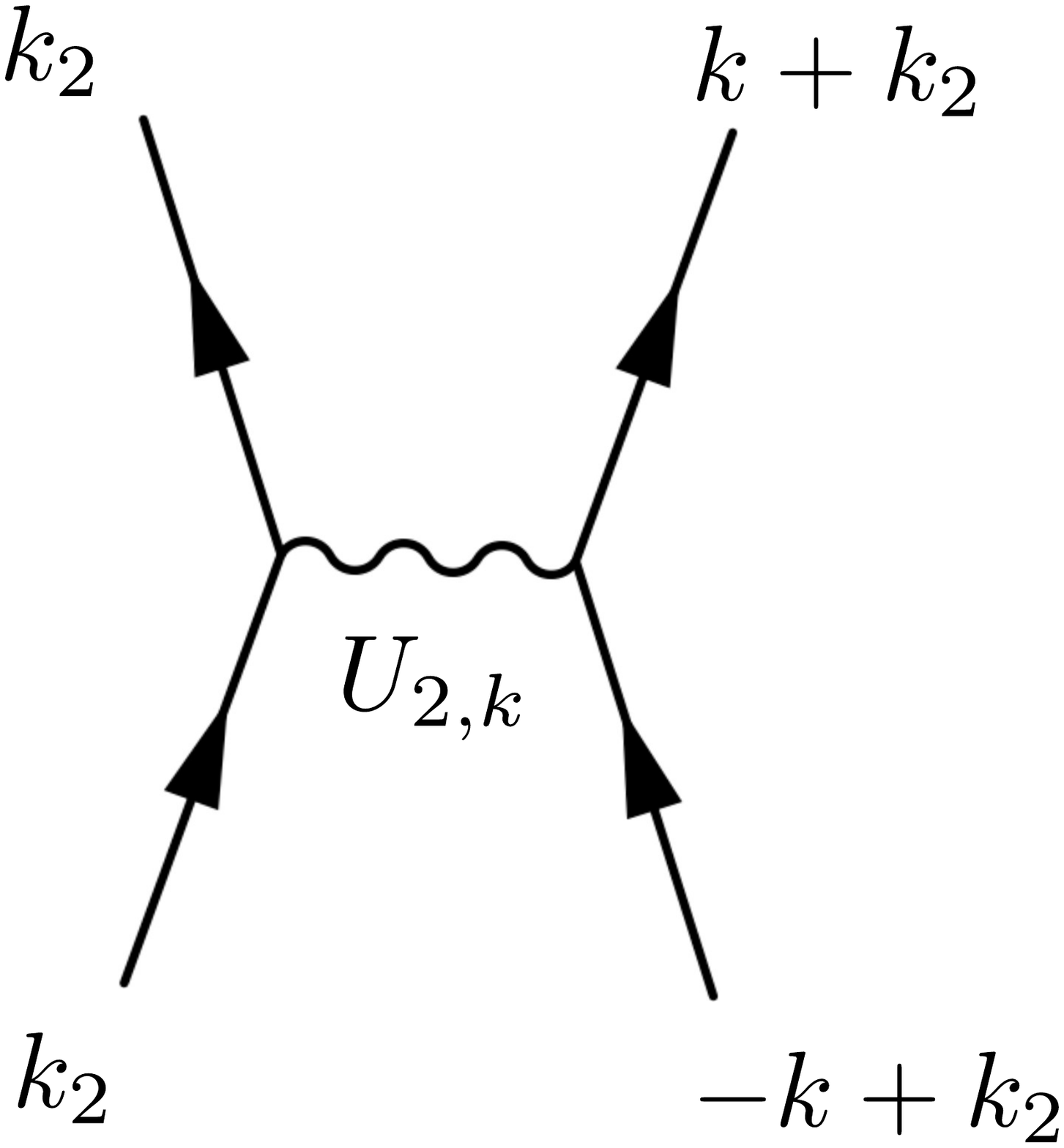}
\includegraphics[scale=0.12]{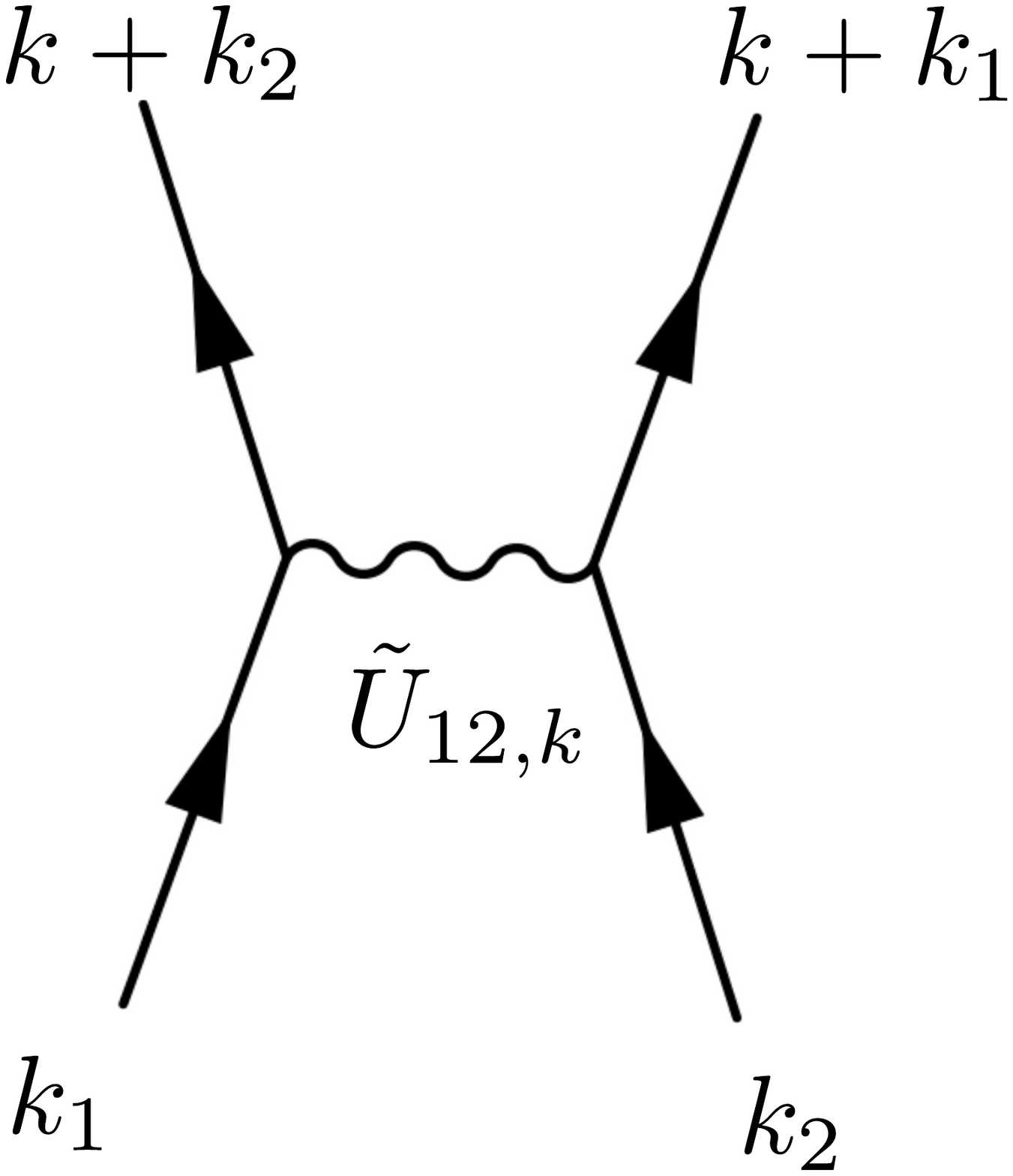}
\includegraphics[scale=0.12]{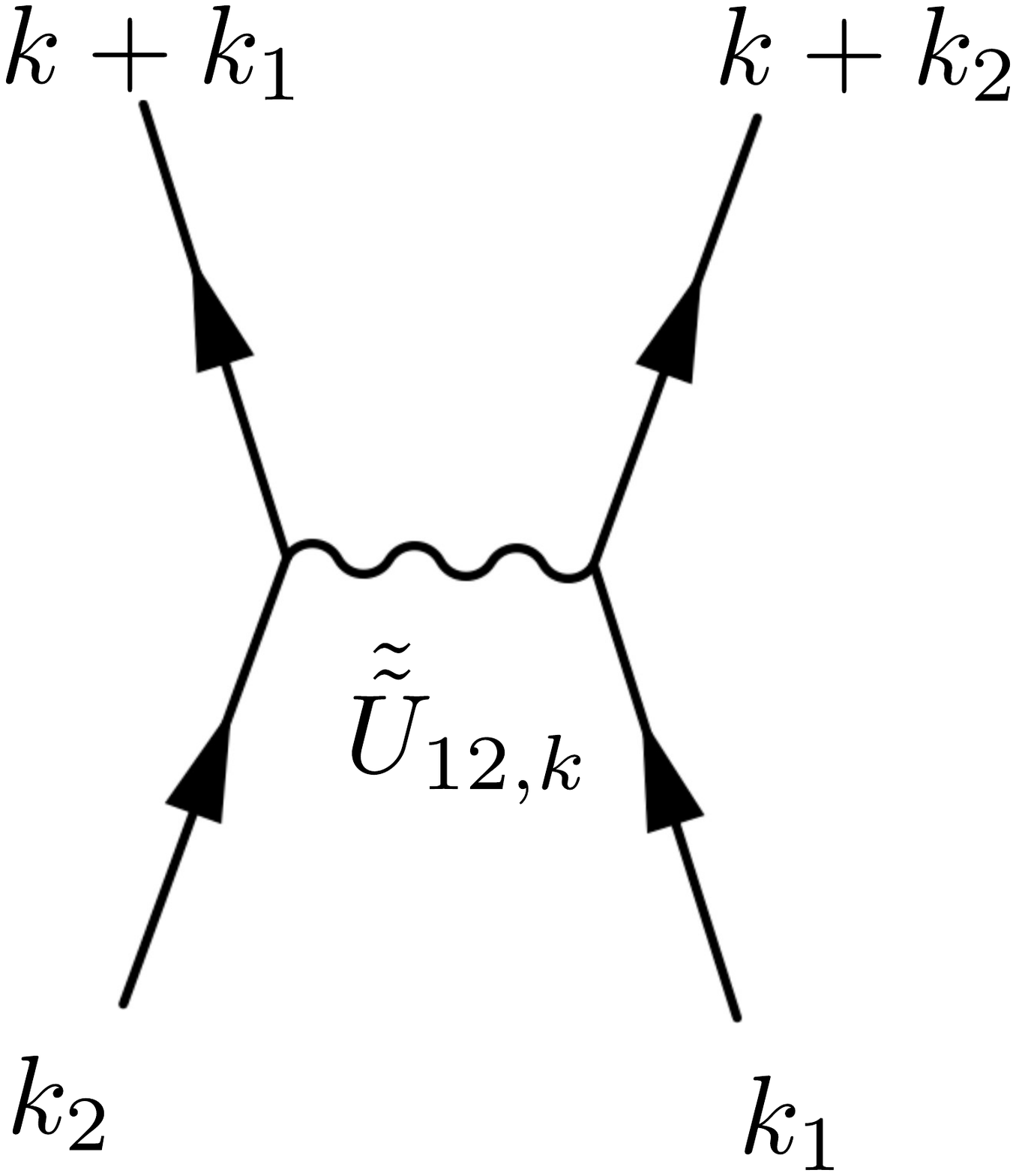}
\caption{Scheme depicting the main collision channels for the the effective interactions among particles belonging to the lower branch of the single-particle excitation spectrum.}
\label{Feymann}
\end{figure}

\section{Dynamical structure factor of the non-interacting case and full expression in the Bogoliubov approximation}
By taking the ground state as being $|0\rangle=\frac{1}{\sqrt{2}}(|k_1\rangle + |k_2\rangle)$ the dynamical structure factor readily reads 
\begin{align}
S_{\beta,\beta}(q,w)=\frac{1}{2}(\delta(\omega-\omega_{q+k_1})+\delta(\omega-\omega_{q+k_2}))
\end{align} 
with $\omega_k= E_{-}(k)-E_-(k_1)$. We see that it consists of two band that correspond only for $k_1=k_2=0$ (Meissner phase) or $k_1=-k_2=-\pi$.

Using Eq.(\ref{Transfo}) and Eq.(\ref{Dynamical}), the dynamical structure factor in the Bogoliubov approximation reads
\begin{align}
&S_{\beta,\beta}(q,\omega)\nonumber\\
&=\sum_{s\neq 0}|\sum_k u_{k+q}u_{k}\left((\tilde{h}_{s,k+q}^{(1)})^*\tilde{\Psi}^{(0)}_{k,1}-(\tilde{\Psi}_{k+q,1}^{(0)}\tilde{Q}_{s,-k}^{(1)})^*\right)\nonumber\\&+v_{k+q}v_{k}\left((\tilde{h}_{s,k+q}^{(2)})^*\tilde{\Psi}^{(0)}_{k,2}-(\tilde{\Psi}_{k+q,2}^{(0)}\tilde{Q}_{s,-k}^{(2)})^*\right)\nonumber\\
&+u_{k+q}v_{k}\left((\tilde{h}_{s,k+q}^{(1)})^*\tilde{\Psi}^{(0)}_{k,2}-(\tilde{\Psi}_{k+q,1}^{(0)}\tilde{Q}_{s,-k}^{(2)})^*\right)\nonumber\\
&+v_{k+q}u_{k}\left((\tilde{h}_{s,k+q}^{(2)})^*\tilde{\Psi}^{(0)}_{k,1}-(\tilde{\Psi}_{k+q,2}^{(0)}\tilde{Q}_{s,-k}^{(1)})^*\right)|^2\delta(\omega-\omega_s)
\end{align}
where $\tilde{h}_{s,k}$, $\tilde{Q}_{s,k}$ and $\tilde{\Psi}^{(0)}_{k,\alpha}$ are the Fourier transforms of $h_{\nu,l}^{(\alpha)}$,$Q_{\nu,\alpha}^{(\alpha)}$ of the excitation and of condensate wavefunction $\Psi^{(0)}_{l,\alpha}$ respectively.




\end{document}